\UseRawInputEncoding
\documentclass[referee]{raa}            
\usepackage{graphicx,times}             
\usepackage{natbib}
\usepackage{amssymb,amsmath}
\usepackage{epstopdf}
\bibpunct{(}{)}{;}{a}{}{,}

\usepackage[pagebackref=true]{hyperref}
\usepackage{verbatim}
\usepackage{longtable}

\begin{document}

  \title{BSEC method for unveiling open clusters and its application to Gaia DR3: 83 new clusters}

   \volnopage{Vol.0 (20xx) No.0, 000--000}      
   \setcounter{page}{1}          

   \author{Zhong-Mu Li
   	\inst{1}
   	\and Cai-Yan Mao
   	\inst{1}
   }

   \institute{Institute of Astronomy and Information, Dali University, Dali 671003, PR China; {\it zhongmuli@126.com}\\
\vs\no
   {\small Received 20xx month day; accepted 20xx month day}}

\abstract{Open clusters (OCs) are common in the Milky Way, but most of them remain undiscovered. There are numerous techniques, including some machine-learning algorithms, available for the exploration of OCs.
However, each method has its limitations and therefore, different approaches to discovering OCs hold significant value.
We develop a comprehensive approach method to automatically explore the data space and identify potential OC candidates with relatively reliable membership determination.
This approach combines the techniques of HDBSCAN, GMM, and a novel cluster member identification technique, color excess constraint. The new method exhibits efficiency in detecting OCs while ensuring precise determination of cluster memberships. Because the main feature of this technique is to add an extra constraint for the members of cluster candidates using the homogeneity of color excess, comparing to typical blind search codes, it is called Blind Search-Extra Constraint (BSEC) method. It is successfully applied to the Gaia Data Release 3, and 83 new OCs are found , whose CMDs are fitted well. In addition, this study reports 621 new OC candidates with discernible main sequence or red giant branch. It is shown that BSEC technique can discard some false negatives of previous works, which takes about 3 percentage of known clusters.
It shows that as an extra constraint, color excess (or 2-color) constraint is useful for removing fake cluster member stars from the clusters that are identified from the positions and proper motions of stars, and getting more precise CMDs, when differential reddening of member stars of a cluster is not large (e.g., $\Delta$E(G$_{BP}$-G$_{RP}$)\,$<$\,0.5\,mag).
It makes the CMDs of 15 percent clusters clearer (in particular for the region near turnoff) and therefore is helpful for CMD and stellar population studies. Our result suggests
that color excess constraint is more appropriate for clusters with small differential reddening, such as globular clusters or older OCs, and clusters that the distances of member stars can not be determined accurately.
\keywords{Galaxy: stellar content -- (Galaxy:) open clusters and associations: general --  stars: fundamental parameters}
}

   \authorrunning{Li \& Mao}            
   \titlerunning{Color excess constraint for star clusters}  

   \maketitle

%
%
\section{Introduction}           
\label{sect:intro}

Open clusters (OCs) are widely recognized as valuable laboratories for studying stellar evolution and as tracers of the structure of the Milky Way galaxy.
They serve as excellent subjects for investigating stellar evolution because stars with the same metallicity and age locate on an isochrone including different stellar evolutionary stages.
In addition, they provide valuable insights into Galactic structure through the observation and measurement of numerous distant OCs and their respective stellar properties.
Consequently, extensive efforts have been dedicated to searching for OCs using diverse datasets, resulting in a significant increase in the number of identified OCs \citep{2023A&A...673A.114H} following the release of Gaia data \citep{2016A&A...595A...1G,2018A&A...616A...1G,2021A&A...649A...1G,2023A&A...674A...1G} in recent years.
The number of well-known OCs are now more than 3000 (see, e.g., \citealt{2022A&A...661A.118C} and \citealt{2023A&A...673A.114H}).
However, the number of known OCs is still significantly lower than the predicted value ($>$ 10000, \citealt{2023arXiv230612894M}).
Besides the limitation of observational data,
the technique for searching for OCs plays an important role. Notably, following the substantial increase in the size of OC samples, which was achieved by
 \cite{1995ASSL..203..127M} (~1200 OCs), the largest increase to the number of Galactic OCs came with the application of new searching methods.
 For example, the OC number of the catalogue of \cite{2013A&A...558A..53K} (hereafter MWSC) reached 2267.
 The recent rise in OC numbers can be attributed to both Gaia data releases and advancements in algorithms designed for identifying these clusters.
 A lot of studies, including \cite{2018A&A...618A..59C,2019A&A...627A..35C,2020A&A...635A..45C,2022A&A...661A.118C}
(CG sample, including 2838 clusters) \cite{2019ApJS..245...32L}, \cite{2019JKAS...52..145S}, and
\cite{2019A&A...624A.126C}, \cite{2022ApJS..259...19L}, \cite{2021A&A...652A.102H}, \cite{2021RAA....21...93H,2022ApJS..260....8H,2022ApJS..262....7H,2023A&A...673A.114H,2023ApJS..265...12Q}
have reported the discovery of over a few thousand candidate OCs based on Gaia data.
Some famous unsupervised machine learning algorithms, e.g., Density-Based Spatial Clustering of Applications with Noise (DBSCAN) \citep{DBSCAN}, (Hierarchical Density-Based Spatial Clustering of Applications with Noise (HDBSCAN) \citep{HDBSCAN}, Gaussian mixture model (GMM) \citep{GMM}, Unsupervised Photometric Membership Assignment in Stellar Clusters (UPMASK) \citep{UPMASK,pyUPMASK}
and  friend-of-friend (FoF) \citep{2005MNRAS.362..711Y}, were employed for blind searches in the Gaia dataset to effectively identify OCs.
However, none of these algorithms is flawless in terms of both efficiency and precision as highlighted by previous studies (see, e.g., \citealt{2021A&A...646A.104H}).
During testing these algorithms, some known clusters were not detected while fake clusters were occasionally reported.
Each method has its own advantages and disadvantages.
According to some tests, GMM is good for determining the cluster membership in a small region whereas HDBSCAN is more suitable for searching OCs on a large scale \citep{2021A&A...646A.104H}.
DBSCAN  is also effective for identifying OCs with similar density \citep{2019A&A...627A..35C}. FoF seems nice for large-scale blind search, but  it often reports false positives like HDBSCAN does.
If membership probabilities are required then GMM and UPMASK can be utilized accordingly.
In order to develop an efficient and relatively precise method for searching OCs we conducted this study.
The work aims to build a new star cluster search method, Blind Search-Extra Constraint (BSEC),
which makes use the advantages of two kinds of existing methods and a new cluster member constraint.
The structure of this paper is as follows.
in section 2 the design of the new method is introduced.
Next, in section 3  the new method is applied to the data of Gaia DR3 \citep{2021A&A...649A...1G}.
Then in section 4 the crossmatch result and CMDs of new OCs are shown.
Finally in section 5 we conclude this work.


\section{Design of BSEC method}
When design the BSEC method,  the advantages of a few different methods are combined. It comes true by the following steps.
First, an effective algorithm is taken to find out star cluster candidates as many as possible.
HDBSCAN and FoF are ideal such algorithms.
Second, a re-identification algorithm is used to re-identify the OC candidates and remove some fake member stars (i.e., field stars).
This step makes the cluster catalogue and member stars of a cluster more reliable. This is important for other works.
Finally, an Extra Constraint (EC) algorithm is used to constrain the member stars again based on some well-known knowledge of star clusters.
In this work, we used the knowledge that all member stars of a cluster have the similar color excess, besides the similar space coordinates and proper motion.
When stars have the same color excess and their color is not too red, metallicity and age, they will distribute on a fixed curve in a color-color diagram (CCD) and on a curve in a color-magnitude diagram (CMD).
Fig. 1 shows the CCD of all stars in the PARSEC-COLIBRI isochrone database \citep{2017ApJ...835...77M}.
These stars cover large ranges in metallicity (0.0002 $\leq Z \leq$ 0.05) and age (3.98\,Myr $\leq age \leq$ 13.2\,Gyr).
We see clearly that when the colors of stars are not too red, e.g., $G_{\rm BP} - G \leq 3.0$ or $G_{\rm BP} - G_{\rm RP} \leq 2.0$, the color-color relation (CCR) of stars can be described well using a polynomial function, with an uncertainty of about 0.05\,mag. For most redder stars, their CCR can also be described via the same fitting function.
As we see, the fitting CCR is almost independent on the mass, age, and metallicity of stars.
There is a fixed fitting function for a group of stars if they have the same color excess, although the fitting formulae are different for various color excesses.
This is very the case of a star cluster, because the member stars have the same color excess for any fixed color.
We can therefore apply the CCR to give a new constraint on the member stars of a cluster. Stars that distribute near the CCR of a cluster are possibly the members of the cluster,
while those far from the CCR are not.
Any two colors can be used for building the CCR of a cluster.
In this work, we take $(G_{\rm BP} - G)$ and $(G_{\rm BP} - G_{\rm RP})$ to build the CCRs.
For a fixed cluster that is observed by Gaia satellite, the CCR is similar to those shown in Fig. 2.
Note that the observed CCR is different from that of PARSEC-COLIBRI stellar isocrhones, because of the color excesses in two colors.
We can kick out the stars that are obviously distant from the CCR curve, as they are not possible the members of cluster.
This technique is called color excess constraint or 2-color constraint.
It is thought as the first part of extra constraint (EC) of star clusters.
Our experiment shows that 2-color constraint kicks out a lot of fake clusters or stars with large photometric uncertainties.

However, the CCR is affected by differential reddening, which may result in significant error when CCR is used. In order to test how differential reddening affects the CCR, we calculate the CCRs with maximum differential reddening values $\Delta$E(G$_{BP}$-G$_{RP}$) from 0.1 to 1.5\,mag. Fig. 3 shows two examples that take the maximum differential reddening values of 0.5 and 1.0\,mag. In the figure, purple squares are stars with random differential reddening $\Delta$E(G$_{BP}$-G$_{RP}$) lower than 0.5\,mag. The orange line is the fitting relation of these stars, while the blue dashed lines denote an error of 0.05\,mag around the fitting curve. We can see that almost all purple squares are within the range of blue lines, indicating that CCR is affected slightly by the differential reddening. Gray dots are for stars with differential reddening smaller than 1.0\,mag. It shows that the color dispersion is significantly lager than the case of differential reddening of 0.5\,mag.
Table 1 lists the errors of fitting CCRs when taking different values for maximum differential reddening. We observe that the fitting error of CCR increases with differential reddening.
It can be concluded that CCR is an effective method for constraining member stars of star cluster when differential reddening is not large, e.g., $\Delta$E(G$_{BP}$-G$_{RP}$) $<$0.5\,mag. Therefore, it is more suitable for studying globular clusters and OCs older than 0.3\,Gyr because of the rare gas and dust. In other word, the color excess constraint may not be appropriate for star clusters with large differential reddening, except there is obvious difference between the reddennings of cluster member stars and the others.
According to the CMDs of the 83 newly discovered OCs in this work, the color spread of main sequence stars near the turn-off is not larger than about 0.6\,mag (including a spread of 0.1\,mag caused by binaries) or that of red giant stars is lower than 0.5\,mag. It suggests that the differential reddening of these OCs is likely smaller than 0.5\,mag. Thus the CCR can be used for giving an EC on the member stars of these clusters.

\begin{figure*}
 \begin{center}
  \includegraphics[angle=0,width=0.9\textwidth]{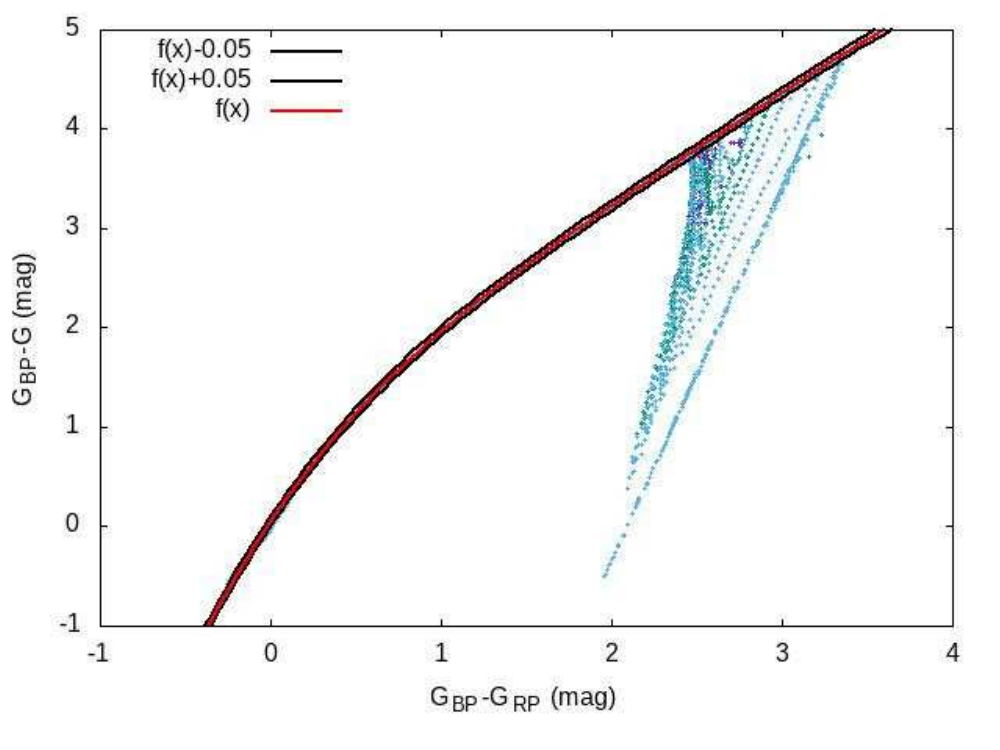}
 \end{center}
 \caption{Color-color diagram of stars in the PARSEC-COLIBRI isochrone database. The metallicity ($Z$) of stars covers the range of 0.0002 to 0.05, while the age of stars covers the range of 3.98\,Myr to 13.2\,Gyr. Points denote stars in the isochrone database. Red line shows the fitting function f(x) = 0.0456 + 2.5451x - 0.8290x$^2$ + 0.2193x$^3$ - 0.0220x$^4$, while black lines show an uncertainty of 0.05\,mag.
}
\end{figure*}

\begin{table}[thp] 
 \footnotesize
 \caption{The errors of best fitting CCRs when taking different values of maximum differential reddening of stars. $\Delta$E(G$_{BP}$-G$_{RP}$)$_{\rm max}$ is for the value of maximum differential reddening.}
 \begin{center}
  \begin{tabular}{ll}
\hline
  $\Delta$E(G$_{BP}$-G$_{RP}$)$_{\rm max}$ & error \\
  $[mag]$ & $[mag]$ \\
\hline
  0.1  & 0.01  \\
  0.2  & 0.01  \\
  0.3  & 0.02  \\
  0.4  & 0.03  \\
  0.5  & 0.05  \\
  0.6  & 0.07  \\
  0.7  & 0.10  \\
  0.8  & 0.12  \\
  0.9  & 0.14  \\
  1.0  & 0.17  \\
  1.1  & 0.20  \\
  1.2  & 0.25  \\
  1.3  & 0.30  \\
  1.4  & 0.34  \\
  1.5  & 0.40  \\
\hline
  \end{tabular}
 \end{center}
\end{table}

\section{Application of BSEC method to Gaia DR3}
\subsection{Blind search of OC candidates}
\subsubsection{Data}
This work uses the data of the latest release of Gaia, i.e., Gaia DR3 \citep{2023A&A...674A...1G}.
It contains the astrometry and broad-band photometry already published as part of Gaia EDR3 \citep{2021A&A...649A...1G}.
This release supplies more stars and more accurate astrometric and photometric data comparing with the first and second releases (DR1 and DR2) \citep{2016A&A...595A...1G,2018A&A...616A...1G}.
No cuts is applied on the observational data, because we need to find out OC candidates as many as possible.
In total, this catalogue contains relatively accurate astrometry and photometry for more than 1.5 billion sources \citep{2021A&A...649E...1F,2021A&A...649A...1G,2021POBeo.100...75D,2023A&A...674A...1G}.
As is well-known, Gaia data have been widely used in the studies of star clusters.

\subsubsection{Algorithm for blind search}
The famous cluster search algorithm, HDBSCAN, is used for blind search of OC candidates.
HDBSCAN is a clustering algorithm that can be used to find cluster candidates in various densities.
This makes it possible to find some small clusters that contains tens of member stars and suitable for this work.
A minimum cluster size $mcl_{\rm Size}$ is set to 25, rather than the suggested value 10,
because a $mcl_{\rm Size}$ of 10 results to too many cluster candidates, which are much more than previous findings.
The leaf mode rather than the default mode (Excess of Mass, i.e., EoM) is taken, because this mode is more effective.
In addition, two candidates within 0.5$^\circ$ are combined in cluster search because the largest OC can reach a size of about 0.5$^\circ$.
In fact, this method may combined two clusters with different distances incorrectly, so we will check the results with parallax distribution later.
Because this is a test, we take each original data file of Gaia DR3 as a grid for blind search.
Although it is somewhat rough, it will not affect the final result obviously, as each file covers enough large coordinate space.

\subsubsection{Result}
The blind search of HDBSCAN reports 6908 cluster candidates.
The candidate number is similar to a recent work, \cite{2023A&A...673A.114H}.
However, as pointed by some papers, e.g., \cite{2023A&A...673A.114H}, there are possibly many fake positive identifications in the results of HDBSCAN.
In addition, the CMDs of some candidates seem strange or contains too many stars (more than 1 million) because no cut is applied on the observational data.
This makes the results not enough reliable and it is impossible to use the data for other researches directly.
Therefore, a re-identification based on the HDBSCAN results have to be done.

\subsection{Re-identification of member stars}
\subsubsection{GMM re-identification}
The GMM algorithm is used to re-identify the cluster candidates, because it is powerful for detecting the cluster members and can supply the possibilities of stars.
Here some cuts are applied on the results of HDBSCAN, for each cluster candidate.
In detail, stars are constrained to be brighter than 21\,mag, with parallax between 0 and 7\,mas and proper motions ($\mu_{\alpha}*$ and $\mu_{\delta}$) between -50 and 50\,mas/yr.
In order to make the results more reliable, only stars around the mean values of proper motions, parallax and coordinates are selected for GMM re-identification.
This kicks out most field stars and make the re-identification much faster and effective.
After the re-identification, we can choose member stars of a cluster according to their possibilities.
As examples, Figs. 4 and 5 show the distributions of member stars after HDBSCAN and GMM processes, in the coordinate and CMD spaces.
We can see that the CMDs become closer to the isochrones of stellar populations if taking larger (e.g, 0.5 and 0.9) possibility cuts (examples 2 and 3 of Fig. 5).
In this work, we take a possibility cut of 0.9 to make sure the member stars reliable.
Cluster candidates that have preferable CMDs, which include at least main sequence or red giant branch, are studied here,
because their CMDs can be compared to stellar isochrones (e.g., \citealt{2017ApJ...835...77M}) and therefore the age and (or) metallicity of such candidates can possibly be determined.
These candidates are more likely to be star clusters.
As a result, 5411 cluster candidates with more than 20 member stars are obtained by the GMM re-identification, within which 1166 ones have preferable CMDs.

\begin{figure*}
 \begin{center}
  \includegraphics[angle=0,width=0.48\textwidth]{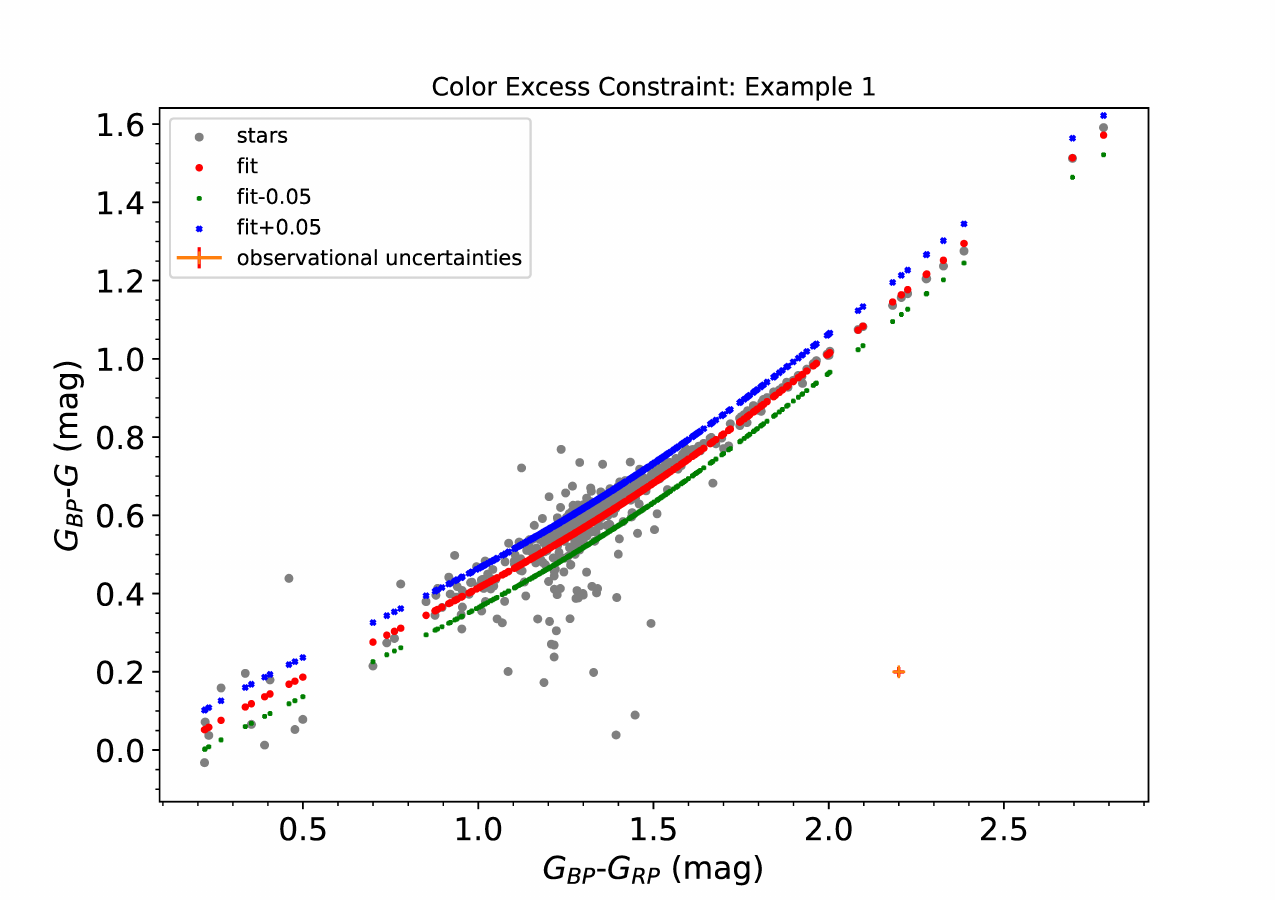}
  \includegraphics[angle=0,width=0.48\textwidth]{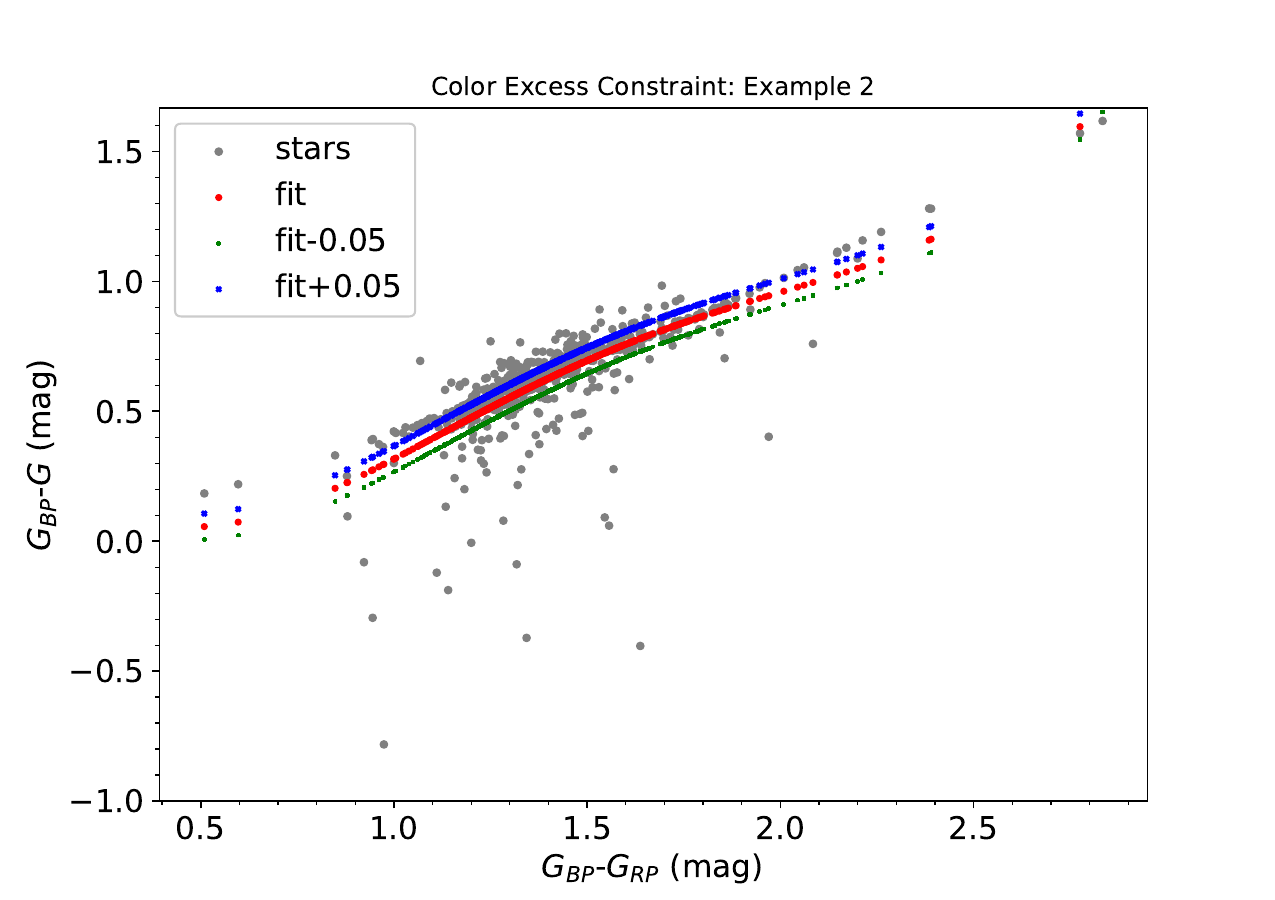}
  \includegraphics[angle=0,width=0.48\textwidth]{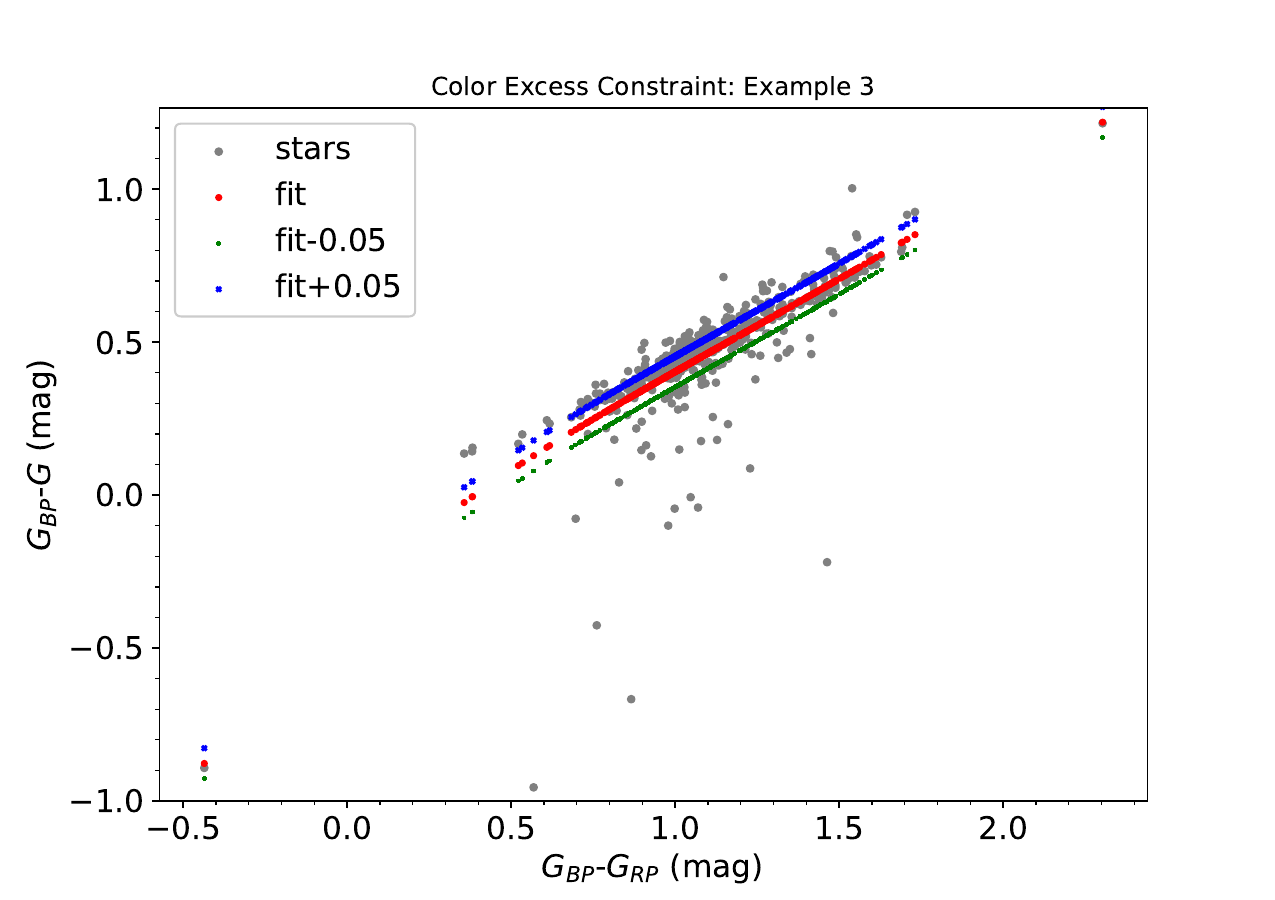}
  \includegraphics[angle=0,width=0.48\textwidth]{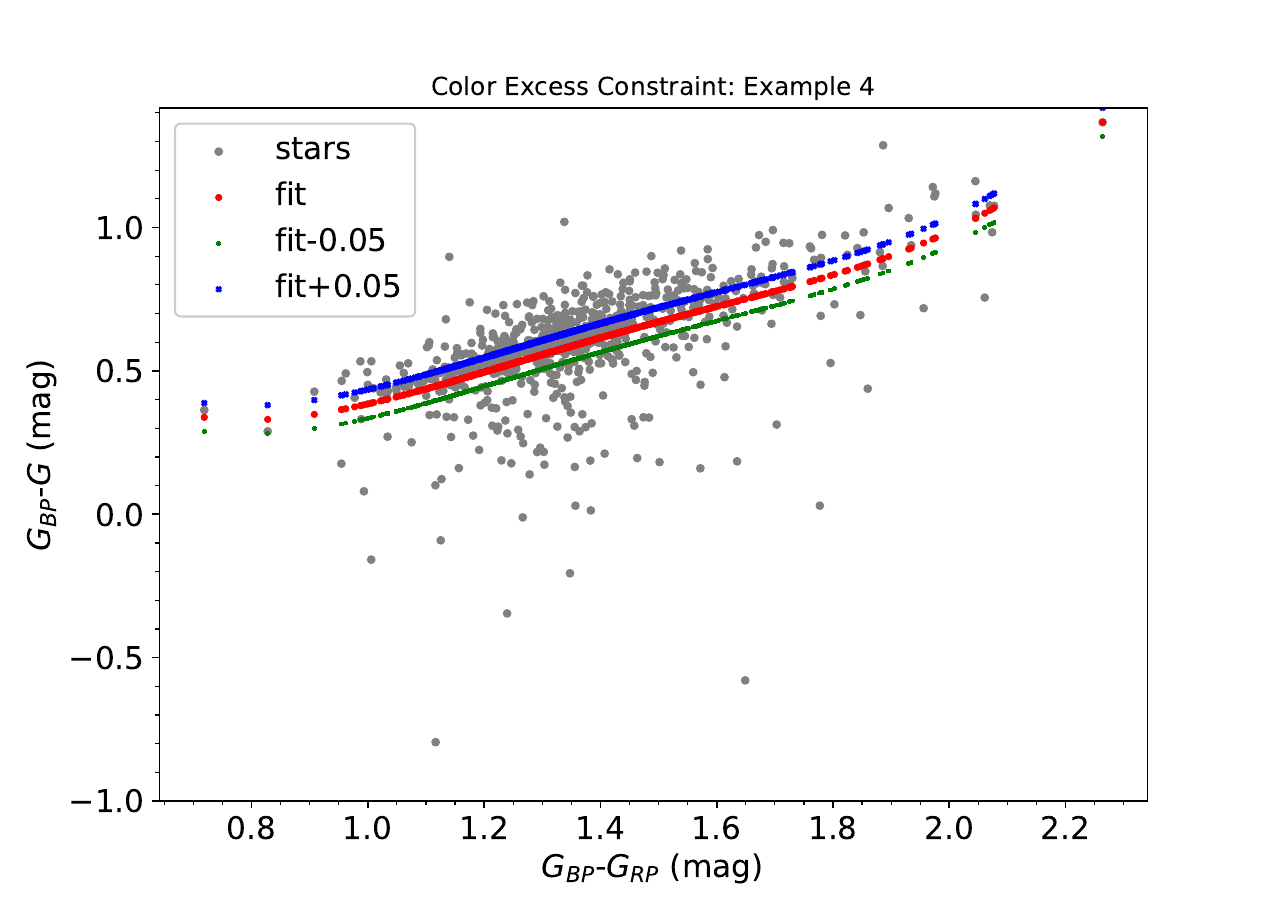}
 \end{center}
 \caption{Color excess constraint (or 2-color constraint) of member stars of clusters, which is based on the color-color relation of stars with the same color excess.
 If there is no observational error, member stars of a cluster should be on a cure like where the red points locate at.
 The gray and red points are for the observed and fitted results.
 The blue and green points indicate the possible range of stars of a cluster if an uncertainty of 0.05\,mag is taken for the colors.
 Only stars distributing between this range are taken as members of a cluster. Error bars denote the color uncertainties at 21\,mag.}
\end{figure*}

\begin{figure*}
 \begin{center}
  \includegraphics[angle=0,width=0.9\textwidth]{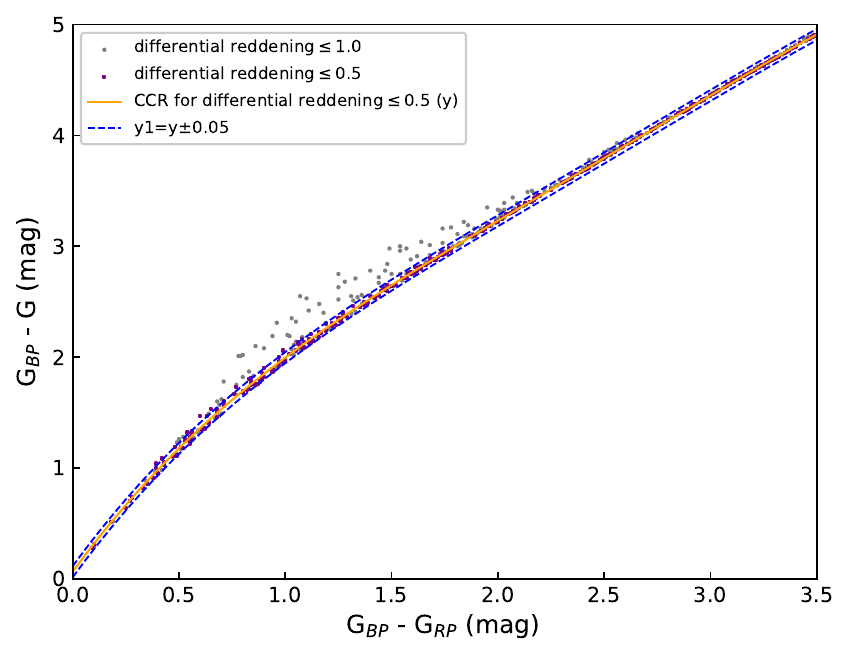}
 \end{center}
 \caption{Example of color dispersions when taking maximum differential reddening $\Delta$E(G$_{BP}$-G$_{RP}$)$_{\rm max}$ of 0.5 and 1.0\,mag. Purple squares are stars with random differential reddening lower than 0.5\,mag. Orange line is the fitting relation of these stars, while blue dashed lines denote an error of 0.05\,mag around the fitting curve. Gray dots are for stars with differential reddening smaller than 1.0\,mag.}
\end{figure*}

\begin{figure}
 \begin{center}
  \includegraphics[angle=0,width=0.48\textwidth]{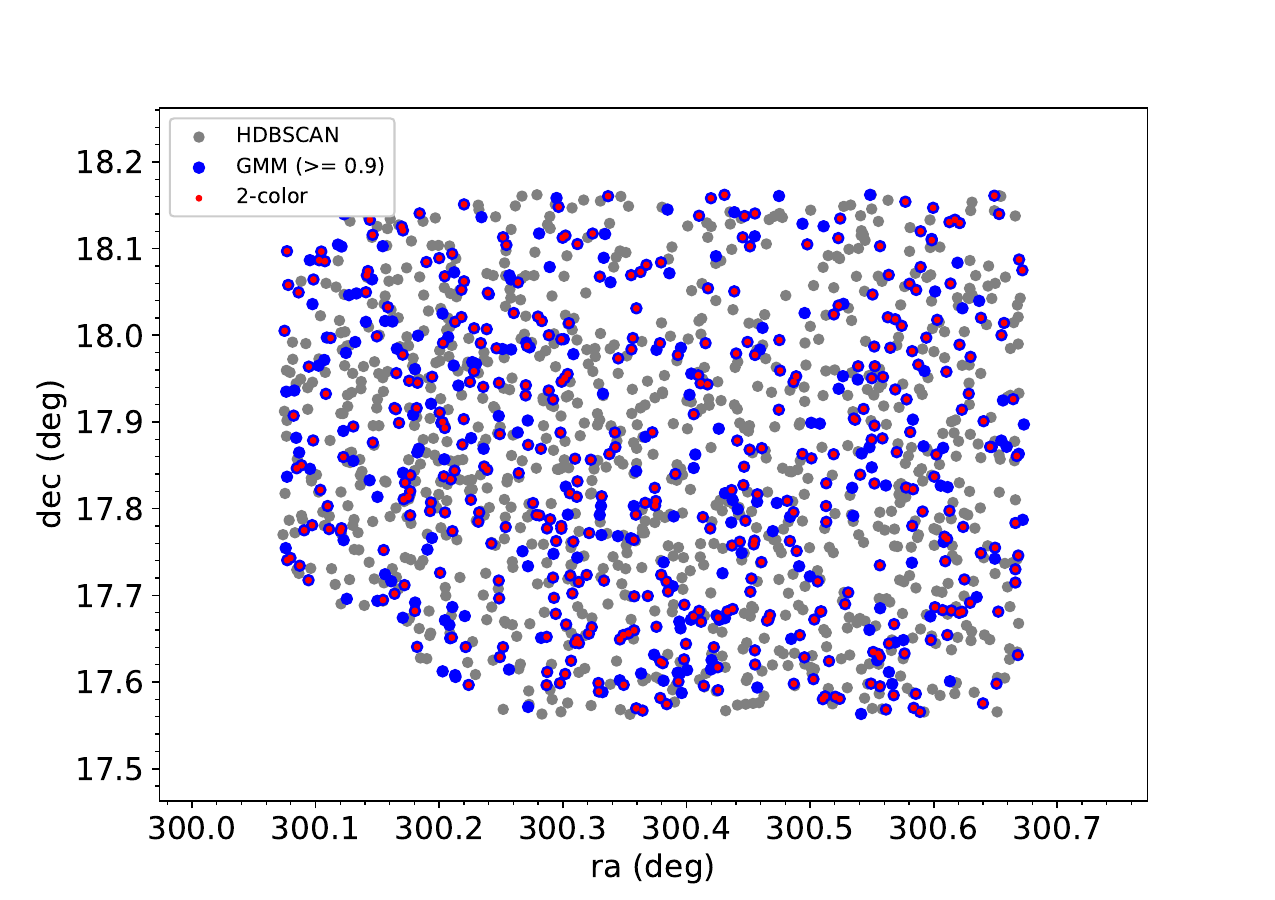}
 \end{center}
 \caption{Distribution of member stars after HDBSCAN, GMM, and 2-color constraint processes.}
\end{figure}

\begin{figure*}
 \begin{center}
  \includegraphics[angle=0,width=0.48\textwidth]{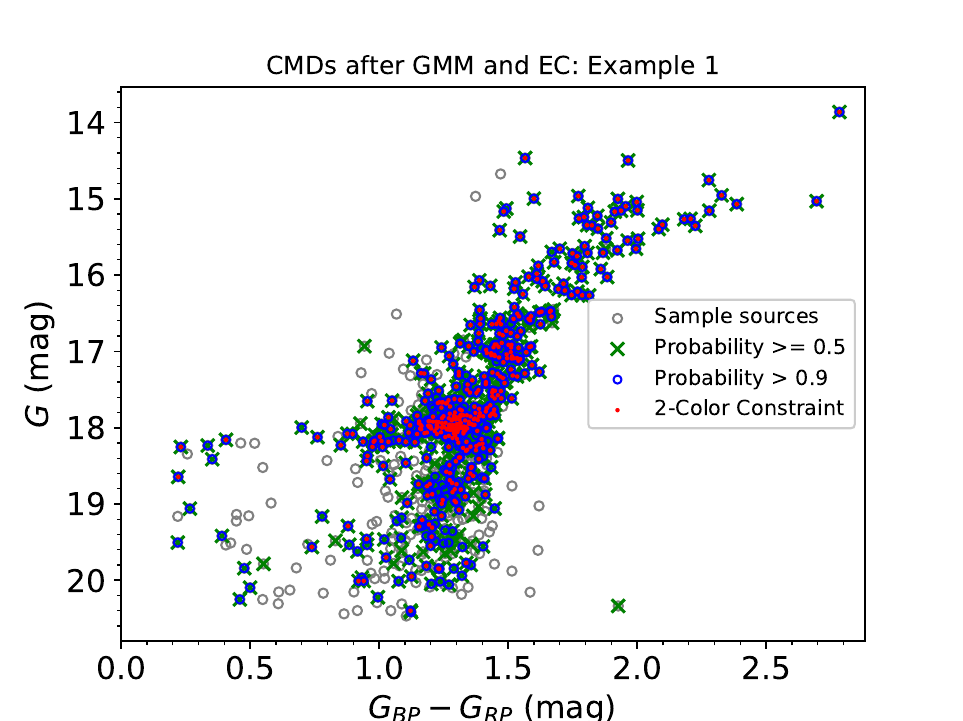}
  \includegraphics[angle=0,width=0.48\textwidth]{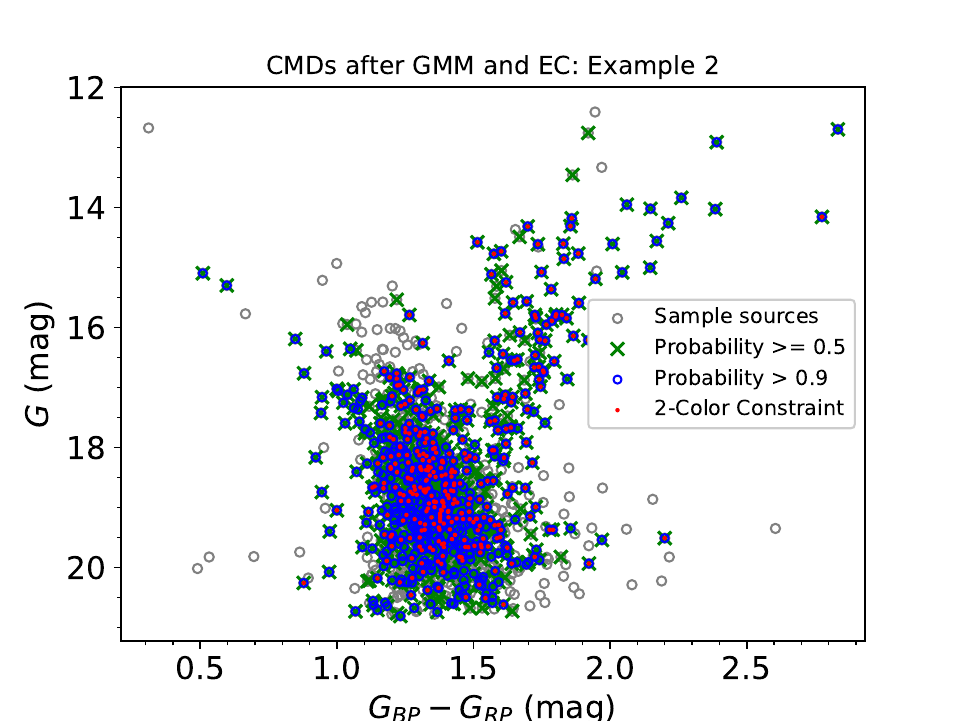}
  \includegraphics[angle=0,width=0.48\textwidth]{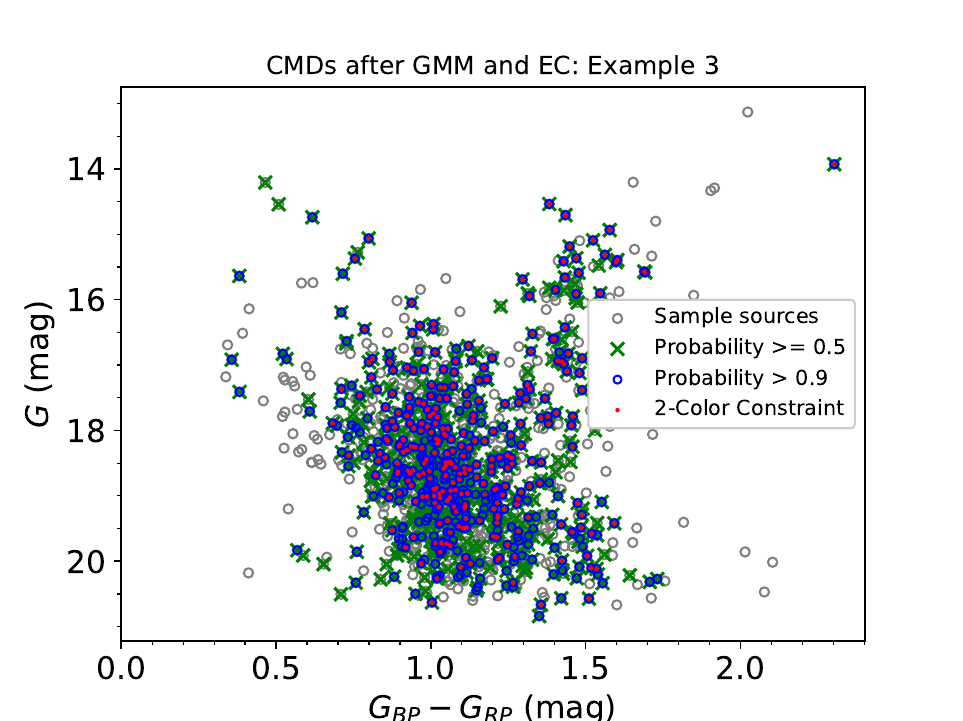}
  \includegraphics[angle=0,width=0.48\textwidth]{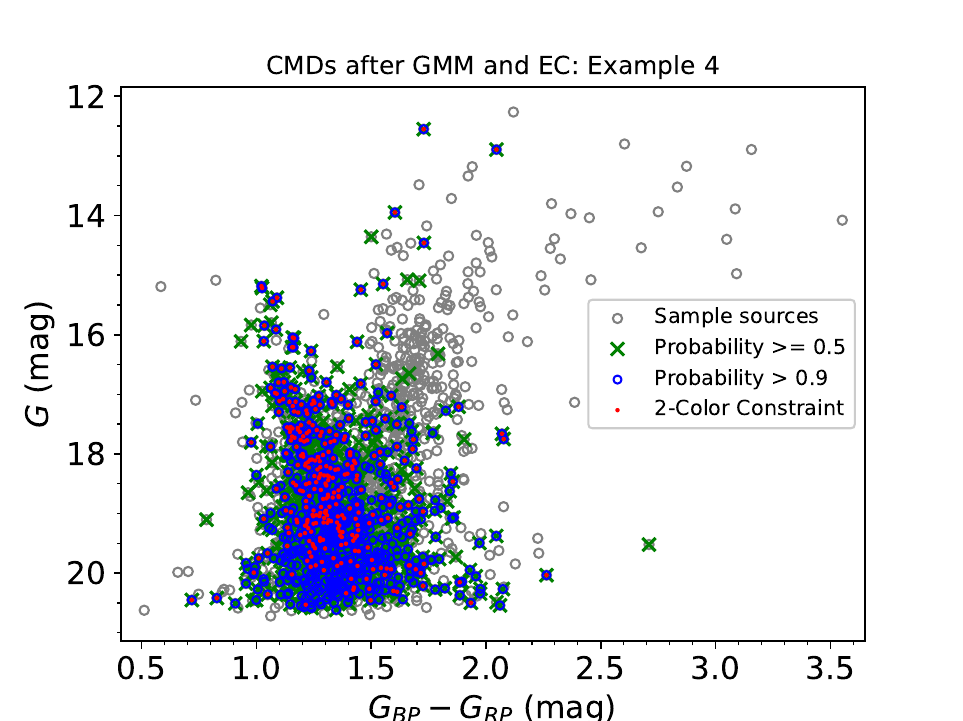}
 \end{center}
 \caption{CMDs of cluster member stars after HDBSCAN (gray open circle), GMM (green cross and blue circle), and 2-color constraint (red point) processes. Stars with two lower limits (0.5 and 0.9) of possibilities are shown for the GMM results.}
\end{figure*}

\subsubsection{Extra constraint}
Because the main structure of CMDs (e.g., main sequence branch, turn off and red giant branch) of some cluster candidates remains unclear and
the astrometric parameters of member stars of some candidates distribute dispersively, an EC constraint is applied to the results of GMM re-identification results.
This process is based on the member stars with GMM possibilities larger than 0.9.
As the first part of extra constraint, the member stars are checked by the color excess information, via the CCR (2-color constraint).
Colors ($G_{\rm BP} - G_{\rm RP}$) and ($G_{\rm BP} - G$) are used for the work.
As shown in Fig. 2, only stars within the possible range are taken as member stars of a cluster.
An uncertainty of 0.05\,mag is taken for stars, because the fitting of the CCR of stars in the PARSEC-COLIBRI isochrone database shows that almost all stars distribute in this range if stars are not too red ($G_{\rm BP} - G \leq 3.0$, see Fig. 1). Note that the four examples shown in Fig. 2 are chosen randomly and therefore are not named.
This constraint eliminates some stars that are possibly not cluster members but it does not change the spatial distribution of stars (Fig. 4).
The CMDs become clearer after 2-color constraint, as we see in Fig. 5.
This will make the CMD fitting of clusters more reliable.
We see that some faint stars are removed from cluster members by the 2-color constraint. This may relate to their relative large photometry uncertainties.
In particular, some fake blue stragglers (on the upper left of turnoff) are kicked out by the 2-color constraint (examples 2 and 3).
This is helpful for both CMD fitting studies and blue straggler studies of star clusters. Note that there is no clear understanding of how the CCR of blue stragglers differ from normal stars, but according to a test based on the data of \cite{2018AJ....156..165C}, blue stragglers obey the same fitting CCR ($U-B$ vs $B-V$) of all stars if the photometric uncertainties and differential reddenning are taken into account.
This constraint is also helpful for getting the real red giant branch (see the upper right of example 2). Because the position of the red giant branch in CMD is sensitive to stellar metallicity, accurate red giant branch is important for determining the metallicities of clusters. Thereby, the color excess constraint is useful for getting reliable metallicity, as it helps to obtain more reliable red giant branch.
The CMDs of 15\% clusters in our work becomes clearer after the 2-color constraint.
However, the CMDs of many cluster candidates are not similar to isochrones of stellar populations yet.
Such candidates may be not real clusters. We therefore check the CMDs of all candidates, as the second part of extra constraint.
The $G$ versus  ($G_{\rm BP} - G_{\rm RP}$) CMD is taken here.
A visual inspection is finally applied to the CMDs of cluster candidates to find out the candidates with clear CMDs.
Besides these, the final clusters are checked by the proper motion dispersions, via the method of \cite{2020A&A...633A..99C} and \cite{2022A&A...660A...4H}.
Equation 1 is used to judge whether a candidate is real cluster or not.
This equation is taken here because newly found clusters have parallax less than about 1\,mas.

\begin{equation}\label{eq1}
     \sqrt{\sigma_{\mu_{\alpha^{*}}}^{2}+\sigma_{\mu_{\delta}}^{2}} \leq 0.5\,{\rm mas}.{\rm yr}^{-1} ~~~~~~~~~~~~~~~~~~~(\varpi<1\,{\rm mas})
\end{equation}

\section{Crossmatch to known clusters and newly found OCs}
When we crossmatch the OC candidates to many catalogs of known open clusters or candidates (e.g., \citealt{
2018A&A...618A..59C,2019A&A...627A..35C,2020A&A...635A..45C,2022A&A...661A.118C,
2001A&A...366..827B,2002A&A...389..871D,2002A&A...384..403R,2003A&A...397..177B,
2003AJ....126.1916P,2003AJ....125.1397C,2004ApJ...602L..21F,2010AstL...36...75G,
2013A&A...558A..53K,2014A&A...568A..51S,
2015MNRAS.448.1930C,2015A&A...581A..39S,
2016MNRAS.455.3126C,2016A&A...593A..95C,2018MNRAS.480.5242K,
2018ApJ...856..152R,2019MNRAS.483.5508F,2019AJ....158..122K,2020A&A...640A...1C,
2021MNRAS.502L..90F,2021A&A...652A.102H,2021RAA....21...93H,
2021A&A...646A.104H,2021MNRAS.504..356D,
2022A&A...660A...4H,
2022A&A...661A.118C,
2022ApJS..259...19L,
2022ApJS..260....8H,
2022ApJS..262....7H,
2023ApJS..265...20C,
2023ApJS..265....3L,2023ApJS..265...12Q}) and globular clusters, 2941 known clusters or candidates (e.g., NGC 5904, NGC 6005, NGC 2420, NGC 6830, NGC 4590, NGC 6584, NGC 6819, NGC 2627, Alessi 6, UBC 285, King 6 and Gulliver 59) are matched within 5411 cluster candidates of this work. Within the 1166 candidates that have preferable CMDs, 621 ones with rough CMDs, which include at least main sequence or red giant branch and have somewhat large disposition, are found to be new.
In particular, 83 ones with good CMDs, which have small disposition and are similar to the isochrones in the PARSEC-COLIBRI database, are found to be new.
Because the distributions of the position, proper motion, color excess, and membership possibility of the member stars of each cluster candidate have been constrained in the previous steps,
the CMD can be employed to impose an additional constraint on these candidates to identify real clusters.
Thus the 83 candidates with good CMDs are more likely to be real clusters. Their stellar populations can be possibly determined from CMDs. We take them as new OCs.
The new clusters are judged according to not only their good CMDs, but also the distributions of the position, proper motion, color excess and membership probability.
Note that if the distance between the centers of a candidate and a known cluster is less than 0.5\,deg, the candidate is thought as a known (or matched) one.
Tables 2 and 3 list the numbers of candidates of this work that are matched to known star clusters or candidates in previous catalogs, for all candidates and those with preferable CMDs respectively.
Fig. 6 shows the distribution of matched and unmatched known clusters and candidates, together with all newly found candidates with preferable CMDs and clusters with good CMDs in the Galactic coordinate system.
We see clearly that the newly found clusters and known clusters are not overlapped.
Table 4 shows the known clusters that are not recovered by this work, while Tables 5 and 6 list the astrometric parameters of newly found OCs with good CMDs.
Meanwhile, Fig. 7 shows the distributions of galactic longitudes and galactic latitudes of matched and unmatched known clusters. We observe that the most matched clusters distribute around $l$ = 0 $\pm$ 75\,deg and $b$ = 0 $\pm$ 10\,deg. It is noted that the number clusters is sensitive to the input parameter of HDBSCAN, mclSize. If a smaller value such as 20 or 10 is taken for searching cluster candidates, much more clusters will be reported. In that case, more known clusters will be matched.

\begin{figure*}
 \begin{center}
  \includegraphics[angle=0,width=0.98\textwidth]{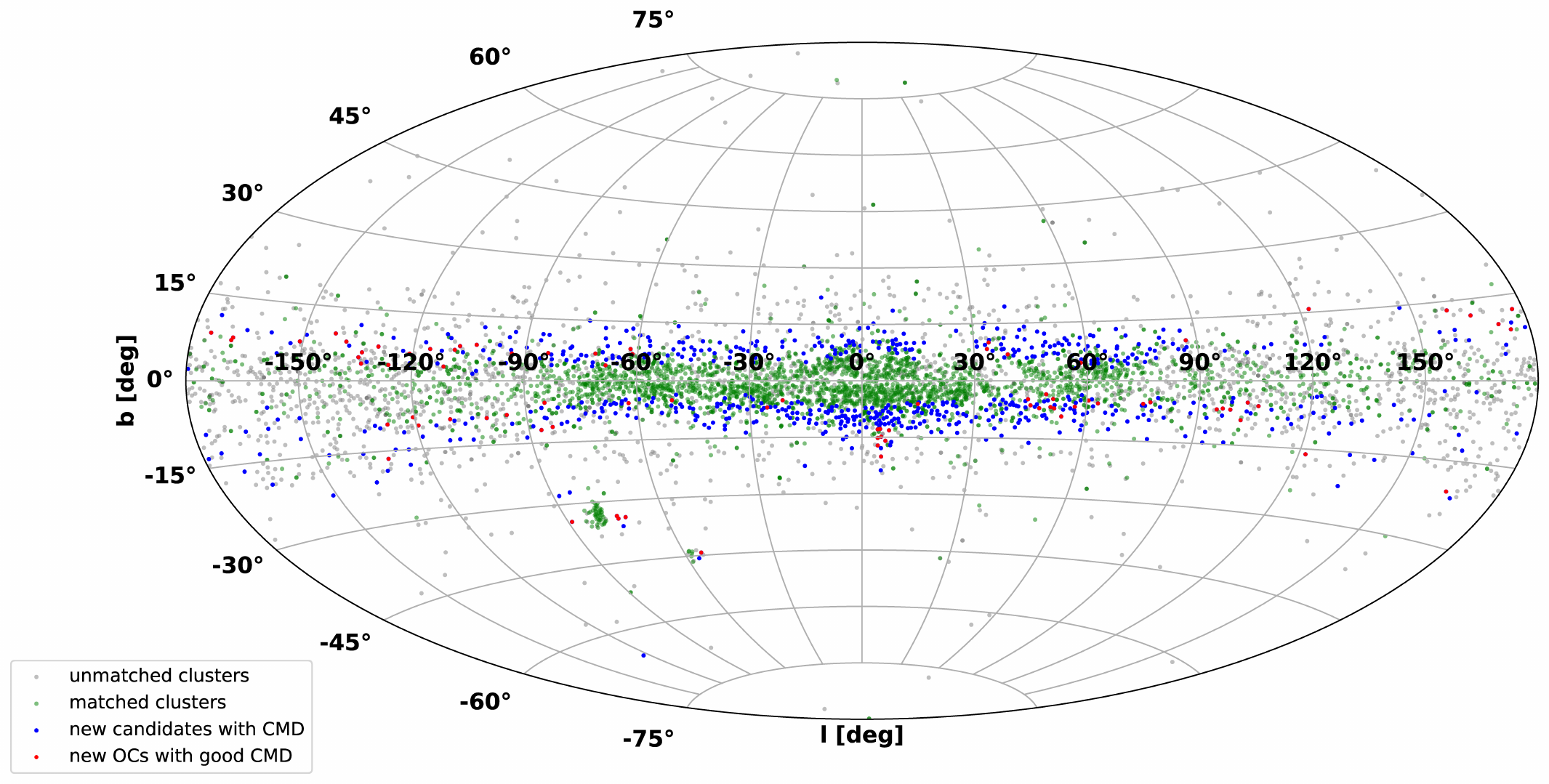}
 \end{center}
 \caption{Distribution of 83 newly found OCs (red) and 621 newly found candidates (blue) in the Galactic coordinate system, together with 2941 matched known clusters and candidates (green) and 2105 unmatched clusters and candidates (grey). The original number of unmatched known clusters is much larger but many clusters are thought as the same one if they are closer than 0.5\,deg.}
\end{figure*}

\begin{figure*}
 \begin{center}
  \includegraphics[angle=0,width=0.5\textwidth]{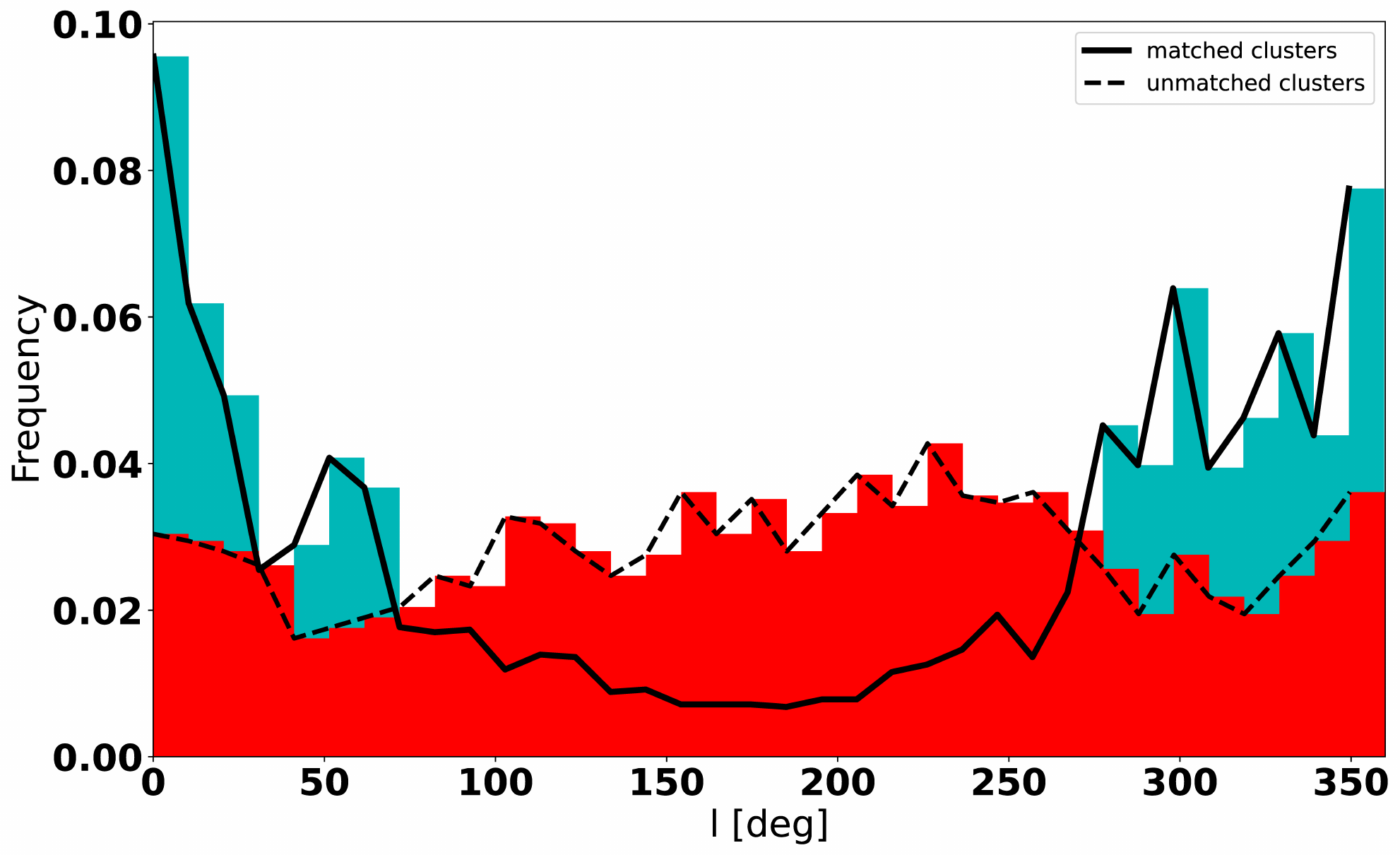}
  \includegraphics[angle=0,width=0.5\textwidth]{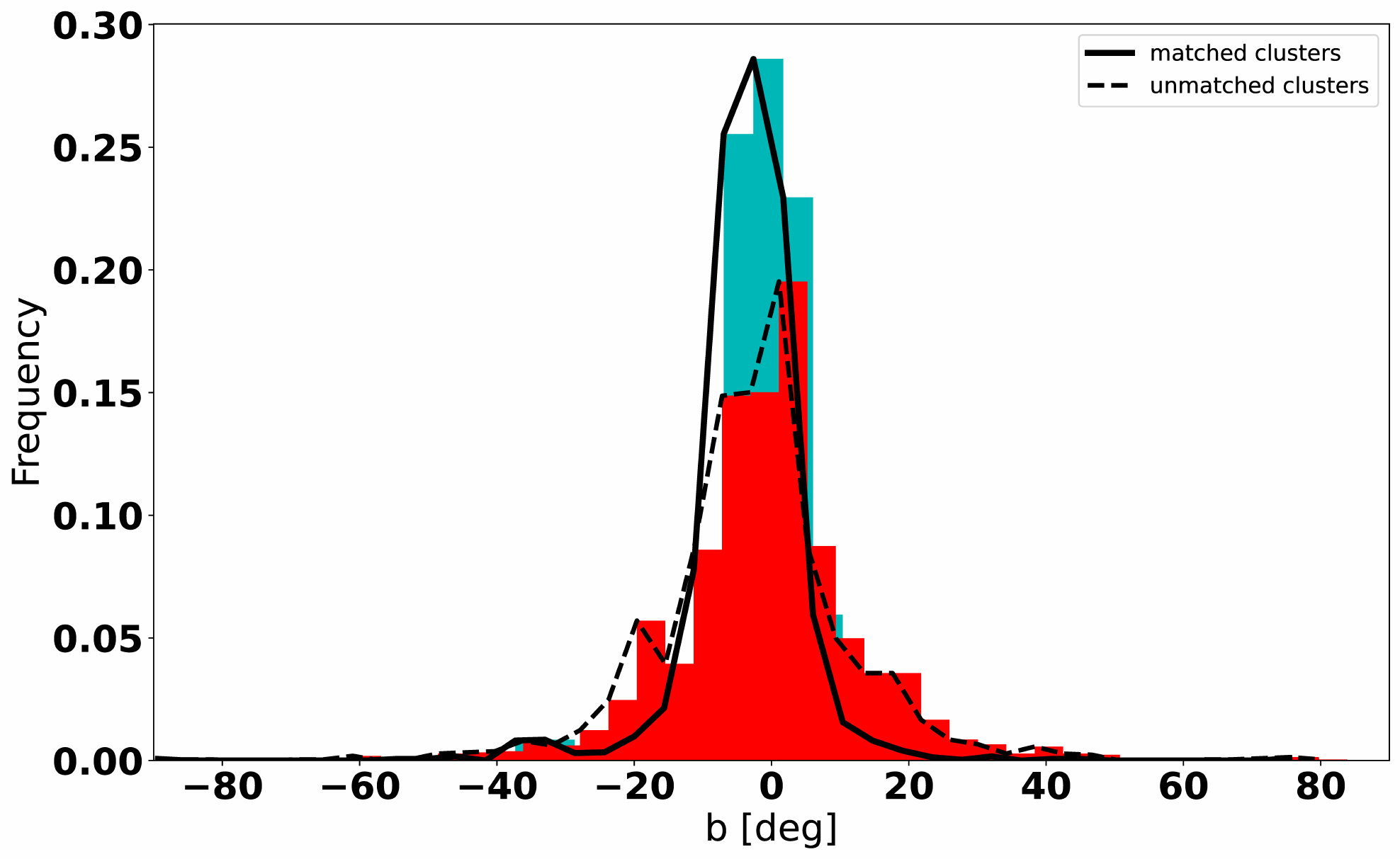}
 \end{center}
 \caption{Distributions of galactic longitudes and galactic latitudes of matched and unmatched known clusters.}
\end{figure*}

\begin{table}[thp] 
 \footnotesize
 \caption{Number of cluster candidates that are crossmatched to different catalogs of known star clusters and candidates. They are shown in green in Fig. 5. Note that a few clusters or candidates in the catalogs of both this and previous works are closer than 0.5\,deg, but they do not change the result obviously.}
 \begin{center}
  \begin{tabular}{lll}
   \hline
Serial Number      &    Catalog   &    Number \\
\hline
1	&Bica et al. (2001)	&1    \\
2	&Reylé \& Robin (2002)	&1    \\
3	&Dias et al. (2002)	&1259 \\
4	&Bica et al. (2003)	&31   \\
5	&Chen et al. (2003)	&1    \\
6	&Porras et al. (2003)	&5    \\
7	&Frinchaboy et al. (2004)	&2    \\
8   &Kharchenko et al. (2013)	&1685 \\
9	&Kharchenko et al. (2005)	&617  \\
10  &Schmeja et al.	(2014)          &4    \\
11  &Scholz et al. (2015)	&10   \\
12  &Camargo et al. (2016a)	&232  \\
13  &Castro-Ginard et al. (2018)	&19   \\
14  &Cantat-Gaudin et al. (2018)	&794  \\
15  &Ryu \& Lee (2018)	&892  \\
16  &Cantat-Gaudin et al. (2019)	&15   \\
17  &Castro-Ginard et al. (2019)	&13   \\
18  &Kounkel \& Covey (2019)	&842  \\
19  &Liu \& Pang (2019) (New OCs)	&50   \\
20  &Ferreira et al. (2019)	&3    \\
21  &Cantat-Gaudin \& Anders (2020)	&936  \\
22  &Castro-Ginard et al. (2020)	&414  \\
23  &Cantat-Gaudin et al. (2020)	&1324 \\
24  &Ferreira et al. (2020)	&28   \\
25  &Hao et al. (2020)	&3    \\
26  &Hunt \& Reffert (2021)	&28   \\
27  &Hao et al. (2021)	&2143 \\
28  &Ferreira et al. (2021)	&48   \\
29  &Dias et al. (2021)	&1167 \\
30  &Hitchcock et al. (2021)	&331  \\
31  &He et al. (2021)	&45   \\
32  &Hao et al. (2022)	&370  \\
33  &Castro-Ginard et al. (2022)	&324  \\
34  &Li et al. (2022)	&2577 \\
35  &He et al. (2022a)	&255  \\
36  &Li \& Mao (2023)   	&191  \\
37  &Chi et al. (2023a)	&18   \\
38  &Chi et al. (2023b)	&931  \\
39  &Milky Way globular clusters               &298  \\
  \hline
  \end{tabular}
 \end{center}
\end{table}

\begin{table}[thp] 
 \footnotesize
 \caption{Similar to Table 1, but for 704 cluster candidates with good CMDs.}
 \begin{center}
  \begin{tabular}{lll}
   \hline
Serial Number      &    Catalog   &    Number \\
\hline
1	&Dias et al. (2002)  	&185\\
2	&Bica et al. (2003)	&5  \\
3	&Kharchenko et al. (2005)	&8  \\
4	&Kharchenko et al. (2013)	&258\\
5	&Scholz et al. (2015)	&1  \\
6	&Camargo et al. (2016a)	&27 \\
7	&Castro-Ginard et al. (2018)	&5  \\
8	&Cantat-Gaudin et al. (2018)	&144\\
9	&Ryu \& Lee (2018)	&42 \\
10	&Castro-Ginard et al. (2019)	&4  \\
11	&Kounkel \& Covey (2019)	&158\\
12	&Liu \& Pang (2019) (New OCs)	&6  \\
13	&Cantat-Gaudin \& Anders (2020)	&177\\
14	&Castro-Ginard et al. (2020)	&61 \\
15	&Cantat-Gaudin et al. (2020)	&236\\
16	&Ferreira et al. (2020)	&3  \\
17	&Hao et al. (2020)	&1  \\
18	&Hunt \& Reffert (2021)	&3  \\
19	&Hao et al. (2021)	&328\\
20	&Dias et al. (2021)	&218\\
21	&He et al. (2021)	&10 \\
22	&Hao et al. (2022)	&35 \\
23	&Castro-Ginard et al. (2022)	&43 \\
24	&Li et al. (2022)	&355\\
25	&He et al. (2022a)	&35 \\
26	&Chi et al. (2023a)	&5  \\
27	&Li \& Mao (2023)  	&31 \\
28	&Chi et al. (2023b)	&195\\
29	&Milky Way globular clusters                	&43 \\
  \hline
  \end{tabular}
 \end{center}
\end{table}

\begin{table}[thp] 
 \footnotesize
 \caption{Example of known clusters that are not recovered by this work. In total, 2105 clusters are not recovered (see grey points in Fig. 5 for their distribution).}
 \begin{center}
  \begin{tabular}{llll}
   \hline
Name      &    l [deg]  &    b [deg] & Catalog \\
\hline
Berkeley 31&206.24&5.134&Cantat-Gaudin \& Anders (2020)\\
Berkeley 59&121.3159&4.6454&Hitchcock et al. (2021)\\
Camargo 506&53.8&0.98&Camargo et al. (2016a)\\
CWNU5&176.325&-10.221&He et al. (2022a)\\
FSR 0555&128.809&8.64&Kharchenko et al. (2013)\\
LISC0115&93.5846&11.8759&Li et al. (2022)\\
LISC3630&300.835&-44.5&Li \& Mao (2023)\\
Majaess 79&205.33&-2.696&Hao et al. (2021)\\
MBM 12&158.7839&-33.9554&Porras et al.(2003)\\
MWSC 5062&250.618&37.05&Schmeja et al. (2014)\\
NGC 7801&114.717&-11.331&Kharchenko et al. (2013)\\
Nor OB5&332.98&1.86&Kharchenko et al. (2005)\\
OC-0038&21.5732&-1.7624&Hao et al. (2022)\\
OC-13&38.7673&12.5237&Hao et al. (2020)\\
PHOC 7&47.2129&4.1091&Hunt \& Reffert (2021)\\
RCW 40&220.79&-1.71&Reyl{\'e} \& Robin (2002)\\
Ryu001&336.7381&-4.9119&Ryu \& Lee (2018)\\
SAI 2&120.5586&13.5431&Glushkova (2010)\\
Stock 21&120.118&-4.812&Hao et al. (2021)\\
UBC1133&89.67&2.7&Castro-Ginard et al. (2022)\\
UBC304&318.8354&-8.3884&Castro-Ginard et al. (2020)\\
UBC4&161.4289&-12.7555&Castro-Ginard et al. (2018), Kharchenko et al.(2013)\\
UBC85&126.0369&-4.8704&Castro-Ginard et al.(2019)\\
UFMG64&6.0804&2.9958&Ferreira et al. (2021)\\
  \hline
  \end{tabular}
 \end{center}
\end{table}

As mentioned above, the clusters with similar $ra$ and $dec$ but different $\varpi$ maybe combined into a cluster incorrectly, so we check our results via the parallax of member stars in each cluster. The parallaxes of stars are divided into 5 or 10 bins, according to the star members of clusters, we find that none of the parallax distributions of 83 new clusters exhibits clearly separate bimodal distribution. It can also be checked via the distribution of member stars in the parallax versus proper motion space (Fig. 8). This suggests that the member stars of each cluster distribute in a concentrated area. Therefore, none of the 83 new clusters includes two physically separated clusters.

In order to understand the properties of the newly discovered OCs, we present the distribution of them in different spaces.
Fig. 8 shows the distribution of some example newly found OCs in the coordinate, proper motion, proper motion versus parallax, and CMD spaces. Here, we plot stars in the parallax versus proper motion space to check whether the stars can be divided into two groups via parallax, and to check the relation between the two parameters.
As we see, most OCs with clear CMDs have clustering features in these spaces.
Figs. 9--11 present the proper motion and parallax distributions of newly found OCs (two samples for those with rough CMDs and good CMDs), together with those of the CG sample \citep{2018A&A...618A..59C,2019A&A...627A..35C,2020A&A...635A..45C,2022A&A...661A.118C}.
We observe that the distributions of proper motion $\mu_{\alpha^{*}}$ and parallax $\varpi$ are different for three samples.
It indicates that the fraction of distant OCs is larger in the sample of newly found OCs comparing to the CG sample.
Thus BSEC is able to find some distant OCs.

\begin{figure*}
 \begin{center}
  \includegraphics[angle=0,width=0.8\textwidth,scale=0.5]{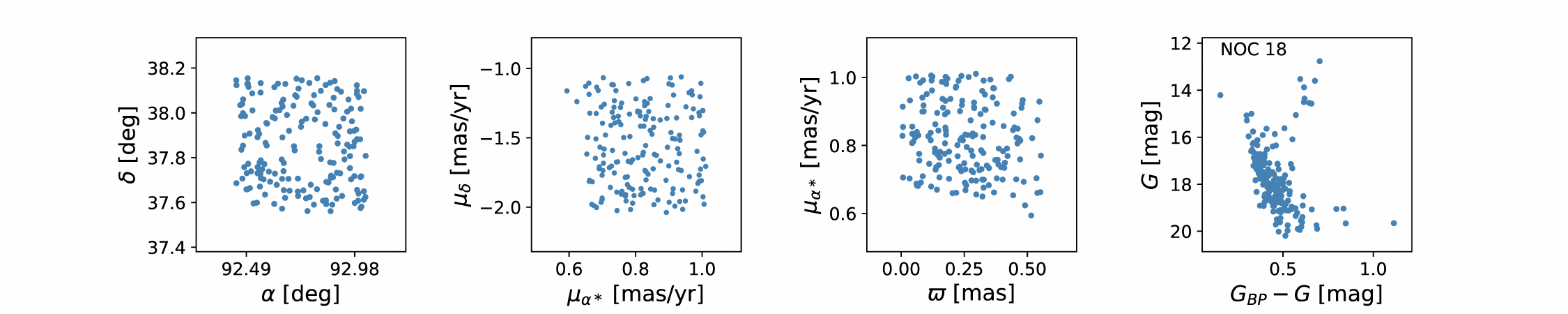}  \includegraphics[angle=0,width=0.8\textwidth,scale=0.5]{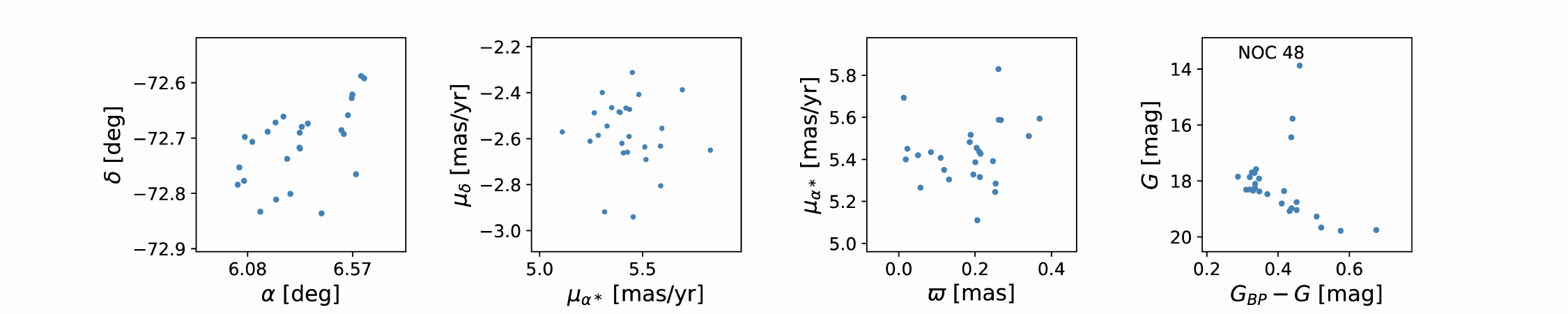}
  \includegraphics[angle=0,width=0.8\textwidth,scale=0.5]{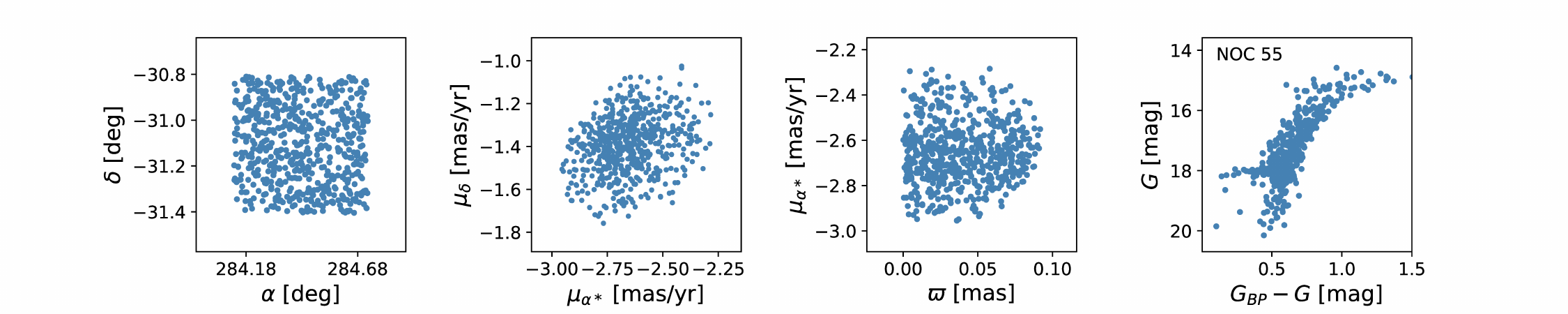}  \includegraphics[angle=0,width=0.8\textwidth,scale=0.5]{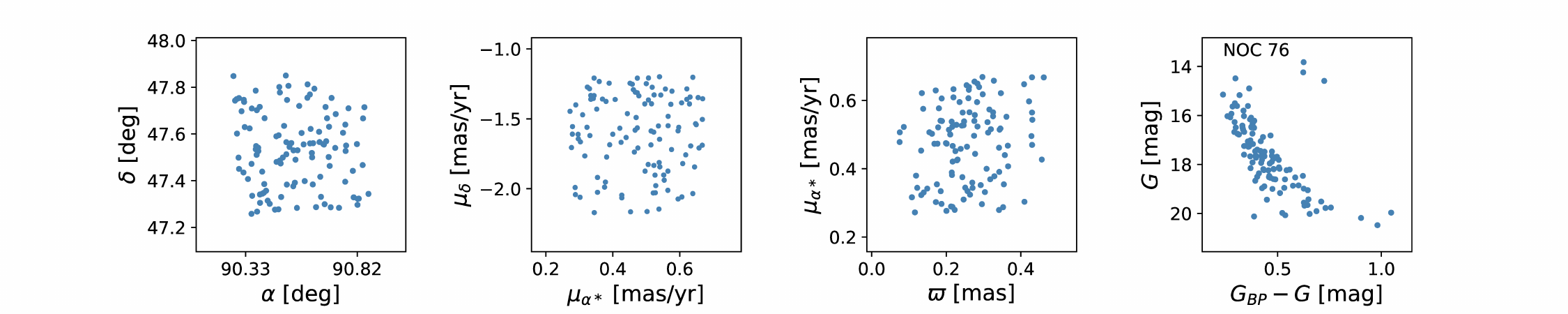}
  \includegraphics[angle=0,width=0.8\textwidth,scale=0.5]{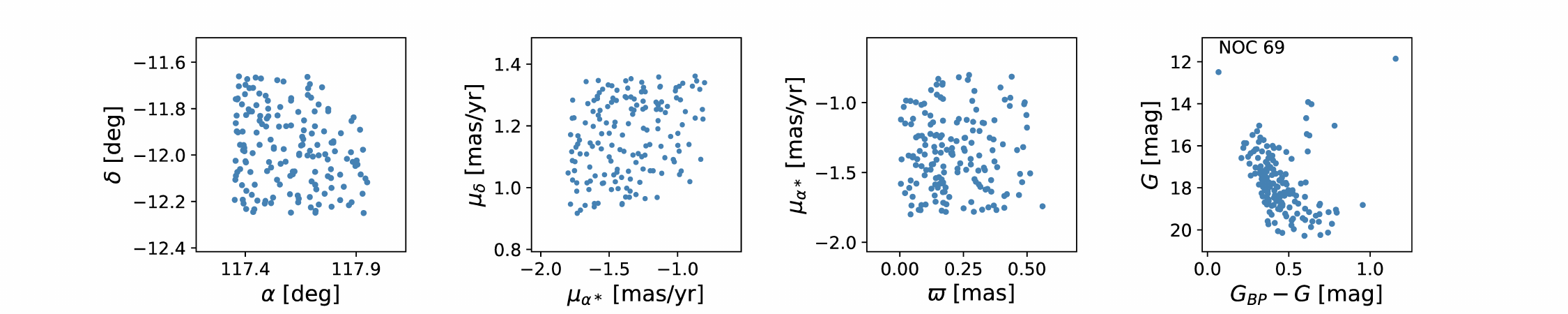}
 \end{center}
 \caption{Distribution of five newly found OCs in the coordinate, proper motion, proper motion versus parallax, and CMD spaces.}
\end{figure*}

\begin{figure}
 \begin{center}
  \includegraphics[angle=0,width=0.5\textwidth]{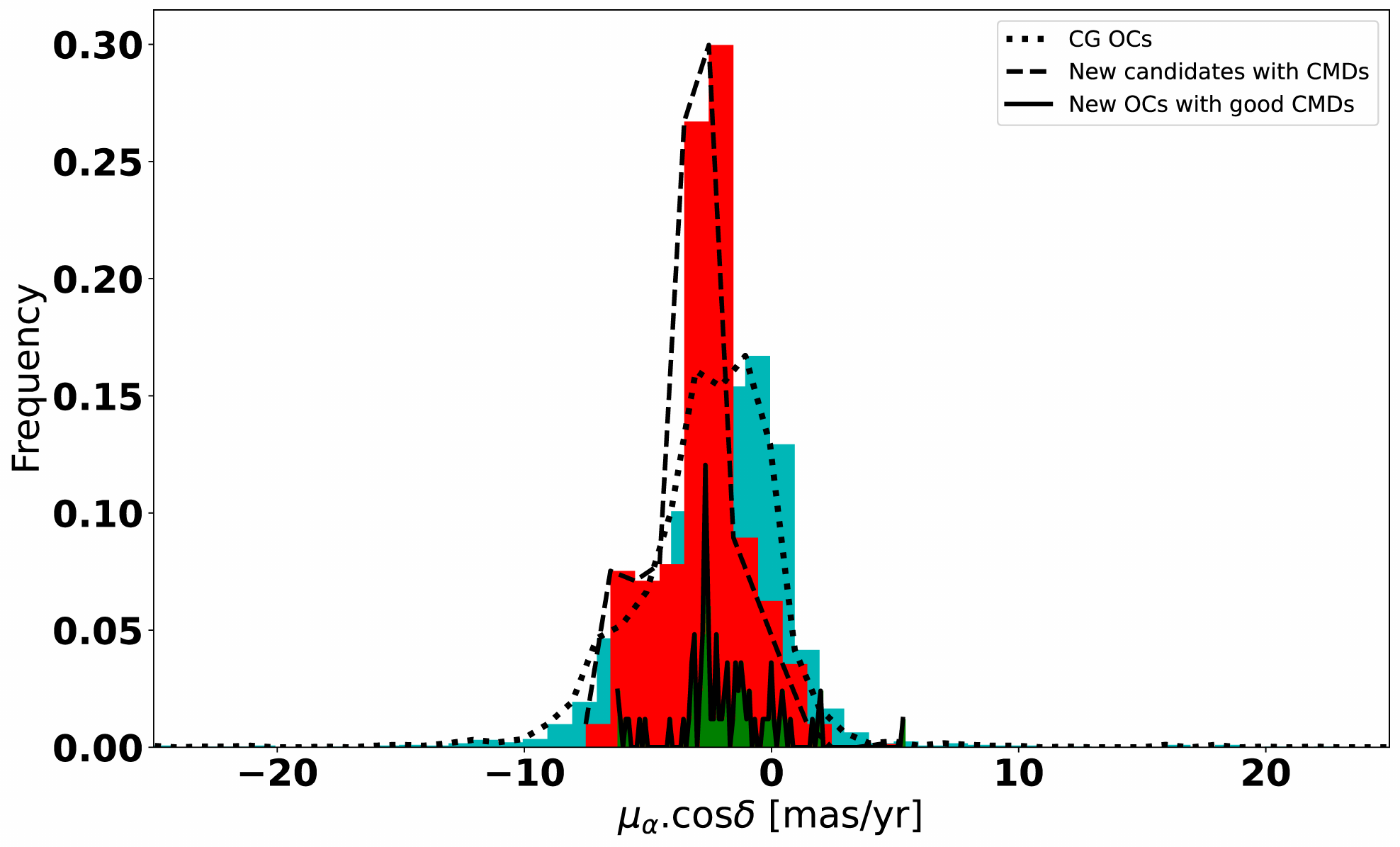}
 \end{center}
 \caption{Comparison of the proper motion distribution ($\mu_{\alpha}$~cos$\delta$) of the newly found OC samples with that of CG sample (dotted line). The dashed and solid lines are for OCs with rough CMDs and good CMDs, respectively.}
\end{figure}

\begin{figure}
 \begin{center}
  \includegraphics[angle=0,width=0.5\textwidth]{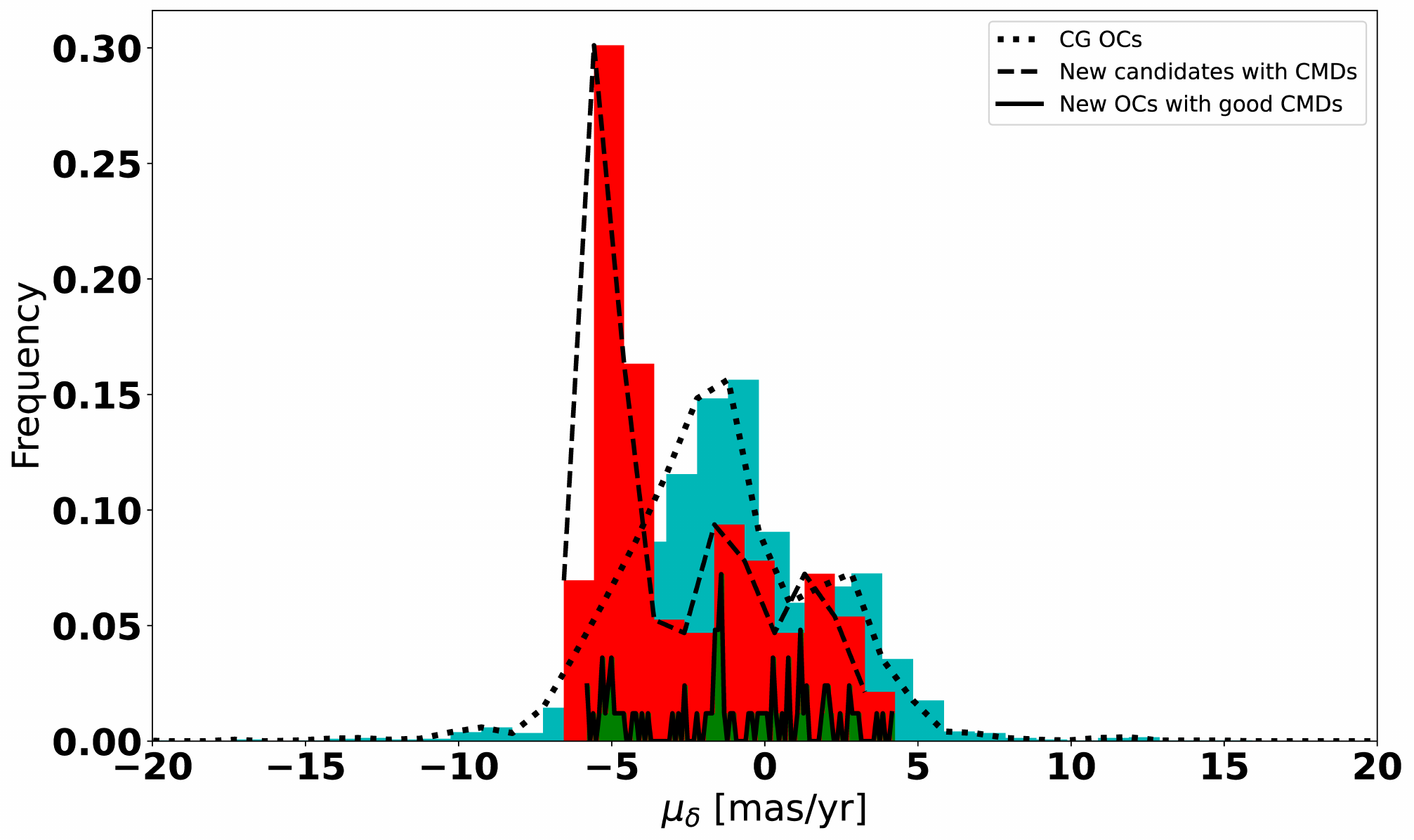}
 \end{center}
 \caption{Similar to Fig. 9, but for another direction ($\mu_{\delta}$).}
\end{figure}

\begin{figure}
 \begin{center}
  \includegraphics[angle=0,width=0.5\textwidth]{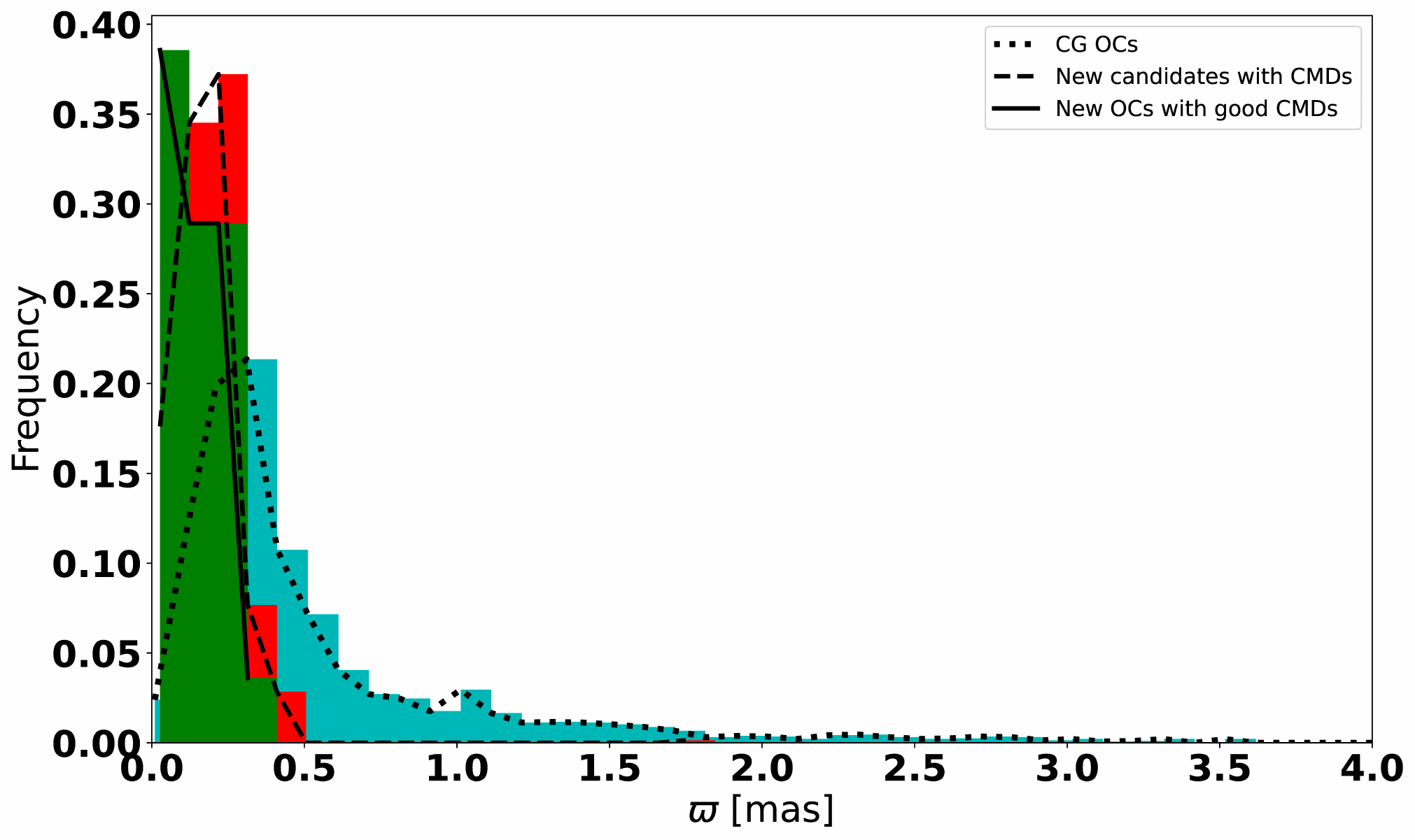}
 \end{center}
 \caption{Comparison of the parallax distributions of the newly found OC sample and that of CG sample.}
\end{figure}

\subsection{CMDs of OCs}
CMD is an important tool for studying OCs. Many fundamental cluster parameters such as color excess, metallicity, age and binary fraction can be determined from CMDs.
Although the CMDs of 83 newly found OCs are relative clear, but only a part of them include both main sequence and red giant branch.
Some clusters have only main sequence or red giant branch. Although their parallaxes suggest that they should be Milky Way clusters, but there are actually obvious uncertainties in the parallaxes.
Figs. 12--14 present the CMDs of the newly found OCs.
The fundamental parameters, such as distance modulus, metallicity and age can be determined by comparing the CMDs to the isochrones of theoretical stellar populations.
The detailed CMD fitting of these clusters will be done and reported in another paper using the stellar population model and CMD fitting code of \cite{2017RAA....17...71L} and \cite{2021ApJS..253...38L}.

\begin{figure*}
 \begin{center}
  \includegraphics[angle=0,width=0.98\textwidth]{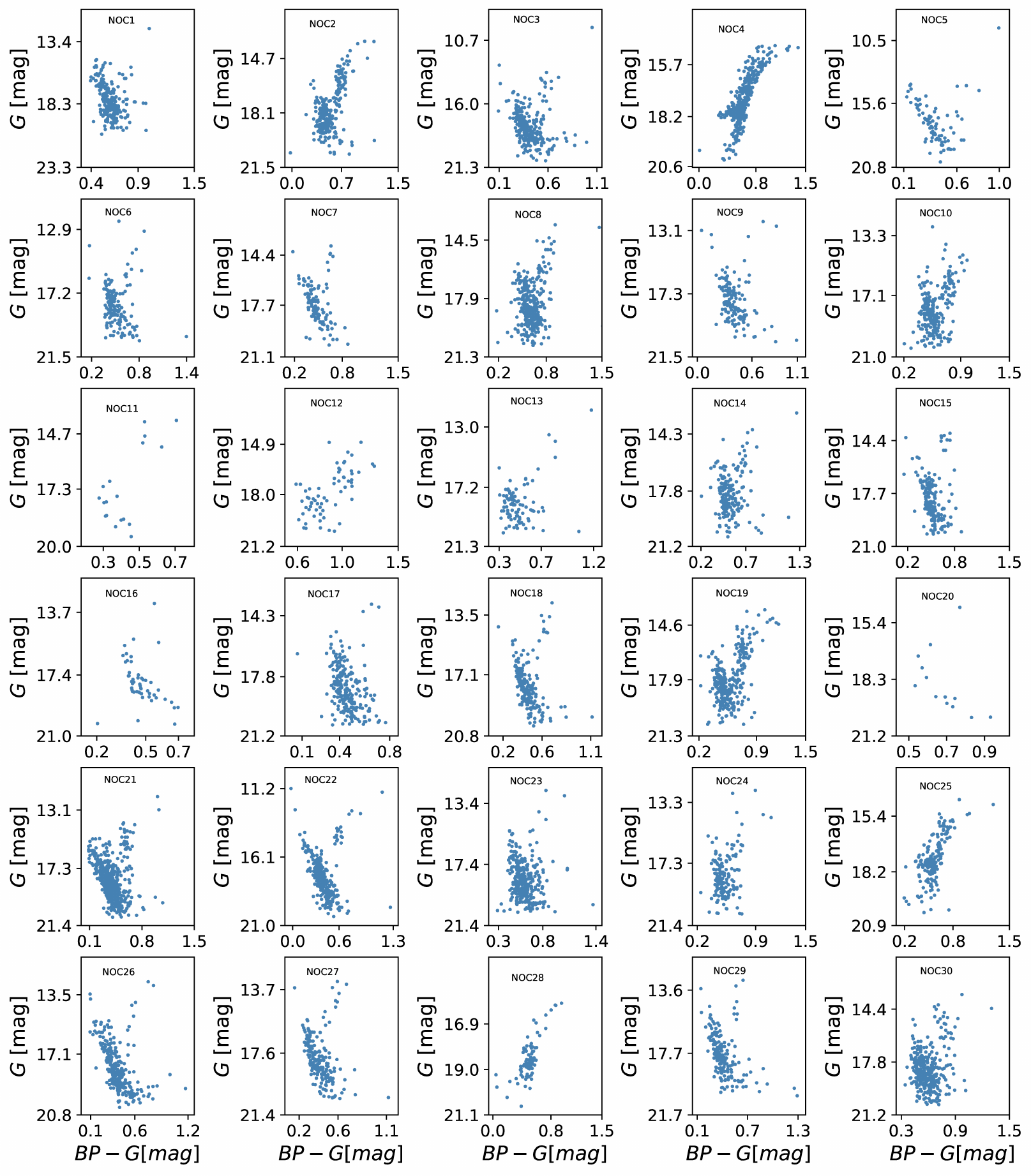}
 \end{center}

 \caption{CMDs of 30 newly found OCs with good CMDs. ``BP'' means ``$G_{\rm BP}$'' magnitude.}
\end{figure*}

\begin{figure*}
 \begin{center}
  \includegraphics[angle=0,width=0.98\textwidth]{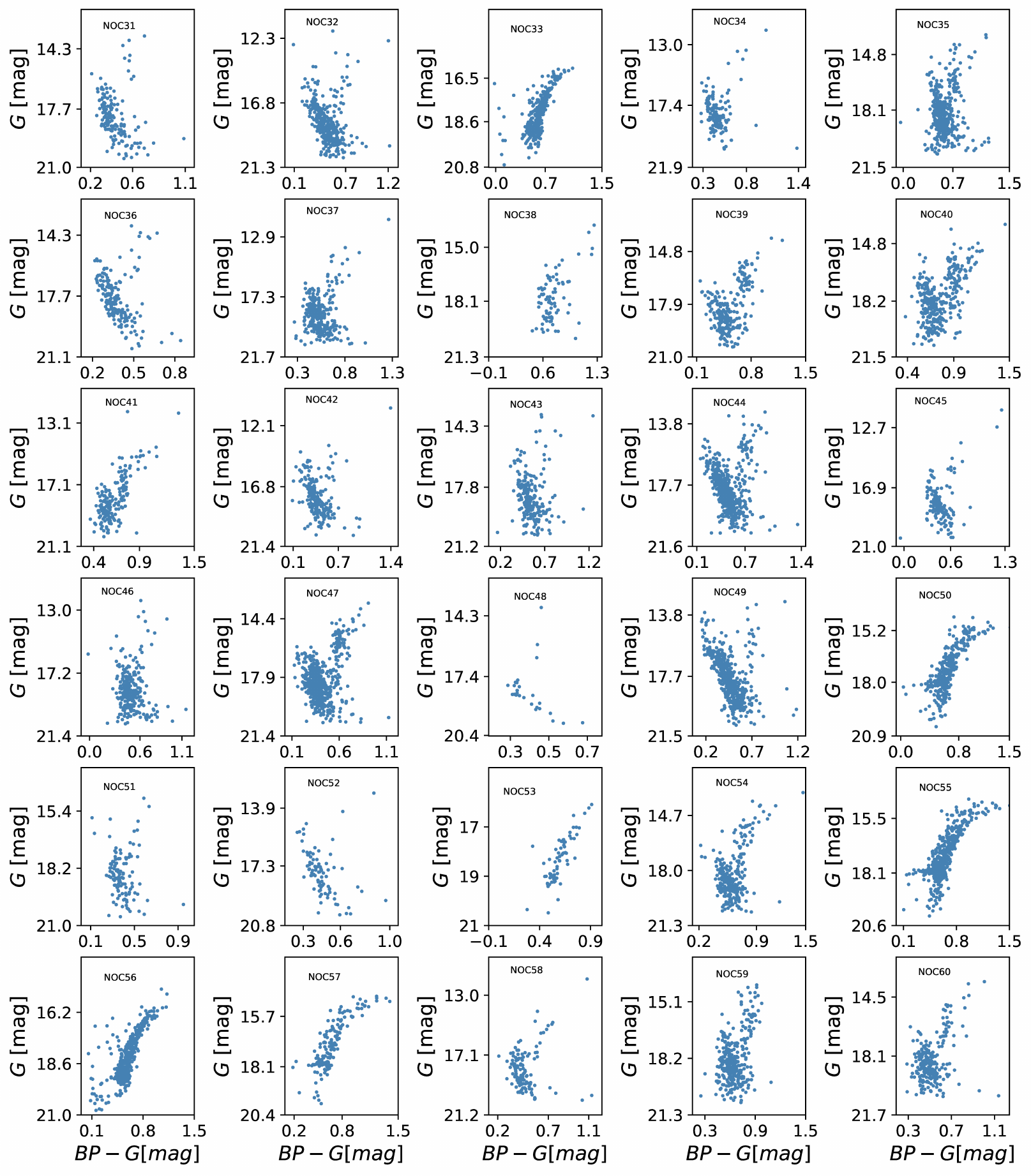}
 \end{center}
 \caption{Similar to Fig. 12, but for 30 other newly found OCs.}
\end{figure*}

\begin{figure*}
 \begin{center}
  \includegraphics[angle=0,width=0.98\textwidth]{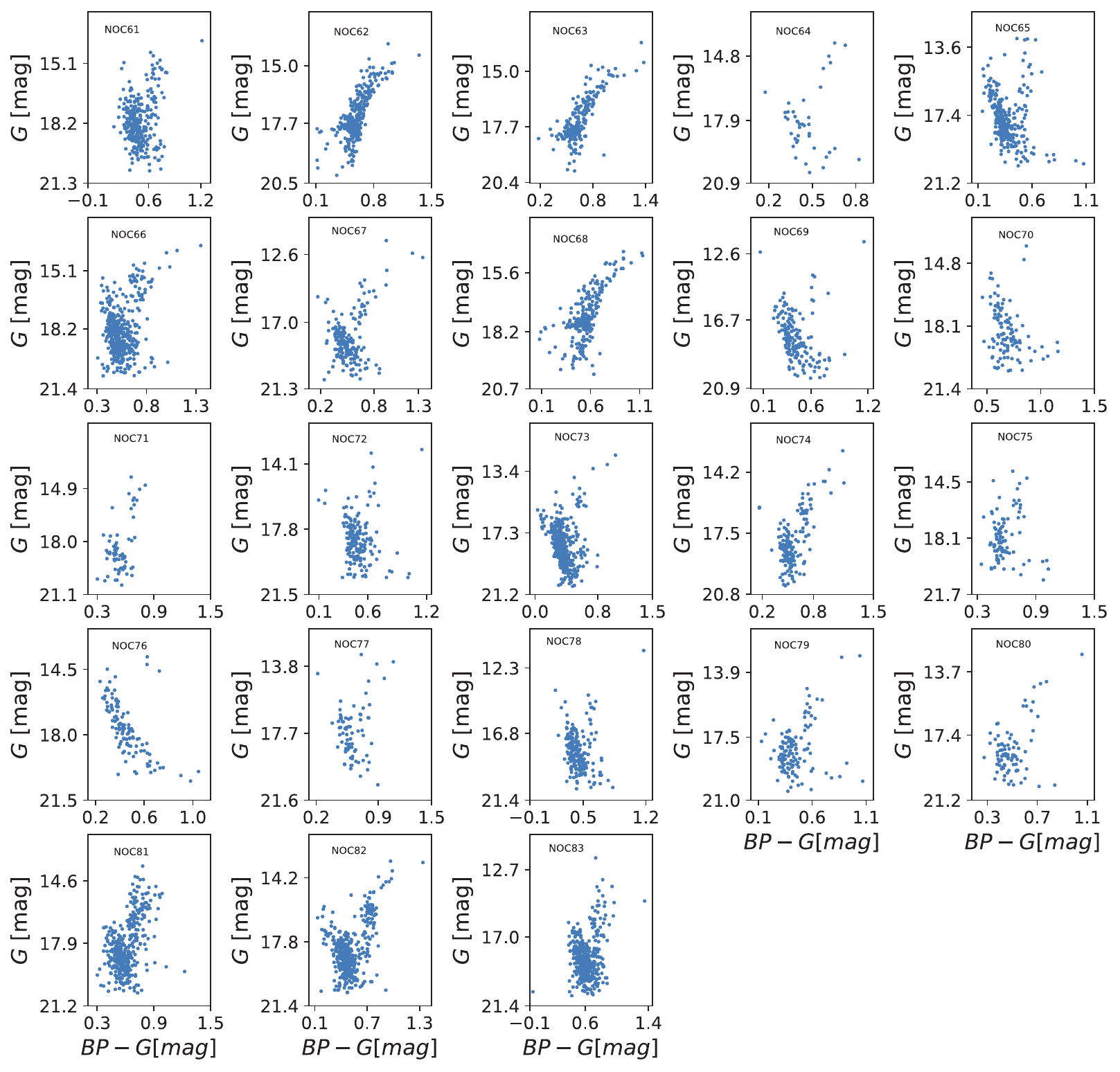}
 \end{center}
 \caption{Similar to Fig. 12, but for 23 other newly found OCs.}
\end{figure*}

\section{Conclusion}
This paper brings forward a new composed method, BSEC, for hunting for star clusters.
The main feature of BSEC is that cluster members are constrained by not only proper motions,
space coordinates, but also the color excess and CMD shape.
The fitted curve of color-color relation is used for constraining the member stars of a cluster when the maximum differential reddening of stars in a cluster is not too large.
This color excess constraint technique improves obviously the membership of clusters and help to obtain clearer CMDs of about 15\% clusters. However, the constraint cannot be independently used for cluster identification. It is better to use it as an extra constraint (EC) of cluster members. Moreover, the EC technique is probably useful for star clusters that the distances of member stars are not determined accurately. It can be used for finding out the background and foreground stars, as their reddenning is usually different from cluster members.
The BSEC method is then applied to the data of Gaia DR3. It finally found 83 new OCs with CMDs similar to the isochrones of stellar populations and 621 new candidates with rough CMDs.
It is worth noting that the space distributions in the $ra$ versus $dec$ space of a few cluster candidates not similar to circles because of the limitation of the observed files used for the study, but this does not affect the conclusion, because this work aims to introduce new methods, BSEC and color excess constraint.
Moreover, we take the number of known real clusters as 3000, BSEC method will discard some false negatives of previous works, which takes about 3\% of known clusters.
The membership and CMDs of these clusters are obtained and their distributions are studied.
Our results show a larger fraction of distant clusters comparing to previous work.
The results can be used in many future studies, in particular the CMD studies.
We can conclude that BSEC is a useful method of cluster identification, which helps for getting more precise CMDs, especially for clusters with small differential reddening.
Although HDBSCAN and GMM blind search algorithms are used in this work, they can be changed to other ones.
In fact, the results are somewhat sensitive to some adjustable parameters, which contributes to some clusters that are not recovered by this work. The method of EC can also be developed in the future.

\begin{table*}[thp] 
 \footnotesize
 \caption{Astrometric parameters and their 1$\sigma$ dispersions of 53 newly found OCs that the CMDs can be fitted well by isochrones.
 NOC ID is the serial number of the new OCs, while $N_{*}$ is the number of member stars.}
 \begin{center}
  \scalebox{0.86}{
  \begin{tabular}{cccccccccccccccccccc}

    \hline
    NOC ID      &     ra   &    dec   &    $\varpi$  &  $\mu_{\alpha}$~cos$\delta$   &  $\mu_{\delta}$     &  $N_{*}$   \\
              &   [deg]  &   [deg]  &     [mas]    &   [mas/yr]                    &  [mas/yr]         &            \\
    \hline
1 &  293.543$\pm$0.159  &  12.677$\pm$0.148  &  0.310$\pm$0.045  &  -2.832$\pm$0.262  &  -5.344$\pm$0.274  & 220 \\
2 &  178.725$\pm$0.175  &  -67.949$\pm$0.178  &  0.081$\pm$0.040  &  -6.249$\pm$0.284  &  1.240$\pm$0.236  & 258 \\
3 &  120.966$\pm$0.170  &  -20.028$\pm$0.178  &  0.196$\pm$0.113  &  -1.386$\pm$0.254  &  1.098$\pm$0.121  & 283 \\
4 &  286.079$\pm$0.162  &  -30.500$\pm$0.166  &  0.042$\pm$0.026  &  -2.690$\pm$0.164  &  -1.460$\pm$0.160  & 446 \\
5 &  99.286$\pm$0.146  &  22.987$\pm$0.148  &  0.264$\pm$0.116  &  -0.241$\pm$0.079  &  -1.243$\pm$0.273  & 83 \\
6 &  330.097$\pm$0.170  &  46.574$\pm$0.154  &  0.224$\pm$0.129  &  -2.532$\pm$0.100  &  -2.949$\pm$0.292  & 153 \\
7 &  92.271$\pm$0.153  &  40.870$\pm$0.173  &  0.256$\pm$0.115  &  0.385$\pm$0.098  &  -1.604$\pm$0.270  & 116 \\
8 &  295.919$\pm$0.167  &  9.461$\pm$0.162  &  0.211$\pm$0.113  &  -3.024$\pm$0.109  &  -5.063$\pm$0.276  & 310 \\
9 &  113.852$\pm$0.170  &  -0.275$\pm$0.174  &  0.298$\pm$0.120  &  -1.182$\pm$0.104  &  -0.199$\pm$0.282  & 136 \\
10 &  295.567$\pm$0.163  &  13.756$\pm$0.137  &  0.179$\pm$0.089  &  -3.051$\pm$0.099  &  -4.980$\pm$0.299  & 245 \\
11 &  57.972$\pm$0.149  &  27.059$\pm$0.123  &  0.103$\pm$0.038  &  0.057$\pm$0.283  &  -0.493$\pm$0.112  & 18 \\
12 &  256.578$\pm$0.096  &  -49.406$\pm$0.121  &  0.122$\pm$0.067  &  -2.453$\pm$0.272  &  -4.299$\pm$0.280  & 68 \\
13 &  130.354$\pm$0.175  &  -60.519$\pm$0.172  &  0.206$\pm$0.106  &  -3.056$\pm$0.108  &  4.252$\pm$0.291  & 90 \\
14 &  297.941$\pm$0.190  &  12.920$\pm$0.124  &  0.205$\pm$0.099  &  -3.187$\pm$0.289  &  -5.016$\pm$0.118  & 187 \\
15 &  140.935$\pm$0.155  &  -41.179$\pm$0.154  &  0.227$\pm$0.130  &  -4.016$\pm$0.097  &  2.980$\pm$0.291  & 170 \\
16 &  88.220$\pm$0.167  &  53.178$\pm$0.150  &  0.214$\pm$0.097  &  0.459$\pm$0.187  &  -1.644$\pm$0.249  & 50 \\
17 &  103.114$\pm$0.161  &  -21.458$\pm$0.165  &  0.313$\pm$0.052  &  -0.045$\pm$0.252  &  1.369$\pm$0.293  & 230 \\
18 &  92.743$\pm$0.181  &  37.855$\pm$0.180  &  0.256$\pm$0.141  &  0.822$\pm$0.106  &  -1.562$\pm$0.282  & 177 \\
19 &  171.677$\pm$0.128  &  -57.142$\pm$0.113  &  0.105$\pm$0.050  &  -5.271$\pm$0.287  &  1.364$\pm$0.278  & 292 \\
20 &  11.047$\pm$0.165  &  78.595$\pm$0.167  &  0.174$\pm$0.063  &  -1.088$\pm$0.228  &  0.909$\pm$0.199  & 13 \\
21 &  107.895$\pm$0.171  &  -27.265$\pm$0.169  &  0.158$\pm$0.099  &  -0.632$\pm$0.239  &  1.977$\pm$0.086  & 545 \\
22 &  113.097$\pm$0.153  &  -8.869$\pm$0.147  &  0.198$\pm$0.107  &  -0.871$\pm$0.095  &  0.808$\pm$0.275  & 294 \\
23 &  292.381$\pm$0.160  &  7.860$\pm$0.175  &  0.354$\pm$0.051  &  -3.128$\pm$0.270  &  -5.741$\pm$0.279  & 314 \\
24 &  184.355$\pm$0.176  &  -54.502$\pm$0.177  &  0.209$\pm$0.122  &  -5.818$\pm$0.086  &  0.337$\pm$0.286  & 158 \\
25 &  290.258$\pm$0.157  &  -33.003$\pm$0.161  &  0.050$\pm$0.031  &  -2.632$\pm$0.188  &  -1.489$\pm$0.157  & 203 \\
26 &  114.366$\pm$0.170  &  -9.054$\pm$0.150  &  0.239$\pm$0.129  &  -1.187$\pm$0.112  &  0.647$\pm$0.252  & 252 \\
27 &  129.249$\pm$0.175  &  -29.224$\pm$0.140  &  0.242$\pm$0.114  &  -2.648$\pm$0.109  &  2.856$\pm$0.284  & 181 \\
28 &  80.373$\pm$0.109  &  -61.864$\pm$0.166  &  0.048$\pm$0.031  &  1.658$\pm$0.172  &  0.341$\pm$0.143  & 88 \\
29 &  114.009$\pm$0.172  &  -3.940$\pm$0.168  &  0.214$\pm$0.121  &  -1.370$\pm$0.108  &  0.478$\pm$0.280  & 175 \\
30 &  310.911$\pm$0.165  &  33.306$\pm$0.166  &  0.228$\pm$0.135  &  -3.266$\pm$0.103  &  -4.561$\pm$0.273  & 345 \\
31 &  130.919$\pm$0.181  &  -29.677$\pm$0.169  &  0.149$\pm$0.082  &  -2.546$\pm$0.148  &  2.513$\pm$0.287  & 157 \\
32 &  115.803$\pm$0.160  &  -16.760$\pm$0.151  &  0.223$\pm$0.126  &  -1.026$\pm$0.100  &  1.262$\pm$0.284  & 316 \\
33 &  74.994$\pm$0.164  &  -73.588$\pm$0.170  &  0.035$\pm$0.022  &  2.066$\pm$0.164  &  0.023$\pm$0.180  & 386 \\
34 &  331.750$\pm$0.194  &  47.690$\pm$0.174  &  0.189$\pm$0.075  &  -2.222$\pm$0.185  &  -2.578$\pm$0.132  & 131 \\
35 &  300.381$\pm$0.167  &  17.852$\pm$0.167  &  0.223$\pm$0.122  &  -2.813$\pm$0.106  &  -4.840$\pm$0.289  & 400 \\
36 &  125.075$\pm$0.173  &  -23.856$\pm$0.166  &  0.115$\pm$0.055  &  -2.122$\pm$0.184  &  2.116$\pm$0.296  & 182 \\
37 &  168.643$\pm$0.179  &  -53.381$\pm$0.174  &  0.222$\pm$0.128  &  -5.727$\pm$0.281  &  1.392$\pm$0.123  & 266 \\
38 &  280.126$\pm$0.145  &  -18.827$\pm$0.156  &  0.261$\pm$0.121  &  -1.527$\pm$0.113  &  -4.148$\pm$0.302  & 81 \\
39 &  123.397$\pm$0.171  &  -49.160$\pm$0.174  &  0.069$\pm$0.036  &  -1.727$\pm$0.157  &  2.782$\pm$0.264  & 243 \\
40 &  290.702$\pm$0.146  &  24.655$\pm$0.168  &  0.162$\pm$0.091  &  -2.533$\pm$0.246  &  -5.288$\pm$0.112  & 345 \\
41 &  279.433$\pm$0.163  &  8.508$\pm$0.131  &  0.076$\pm$0.041  &  -2.157$\pm$0.281  &  -4.766$\pm$0.294  & 182 \\
42 &  99.921$\pm$0.169  &  22.927$\pm$0.175  &  0.217$\pm$0.118  &  0.118$\pm$0.255  &  -0.996$\pm$0.122  & 210 \\
43 &  343.851$\pm$0.172  &  53.177$\pm$0.169  &  0.222$\pm$0.120  &  -1.780$\pm$0.101  &  -1.461$\pm$0.283  & 210 \\
44 &  124.006$\pm$0.169  &  -29.385$\pm$0.164  &  0.117$\pm$0.053  &  -1.853$\pm$0.261  &  2.082$\pm$0.270  & 568 \\
45 &  97.509$\pm$0.176  &  28.688$\pm$0.169  &  0.093$\pm$0.049  &  0.076$\pm$0.265  &  -1.363$\pm$0.263  & 140 \\
46 &  300.823$\pm$0.171  &  13.652$\pm$0.173  &  0.225$\pm$0.122  &  -2.688$\pm$0.110  &  -4.665$\pm$0.283  & 254 \\
47 &  118.493$\pm$0.176  &  -44.822$\pm$0.176  &  0.084$\pm$0.041  &  -1.348$\pm$0.245  &  2.870$\pm$0.276  & 576 \\
48 &  6.326$\pm$0.190  &  -72.710$\pm$0.068  &  0.180$\pm$0.095  &  5.431$\pm$0.147  &  -2.579$\pm$0.148  & 26 \\
49 &  112.340$\pm$0.175  &  -10.282$\pm$0.177  &  0.231$\pm$0.125  &  -1.148$\pm$0.248  &  0.838$\pm$0.109  & 470 \\
50 &  282.081$\pm$0.139  &  -30.605$\pm$0.140  &  0.058$\pm$0.034  &  -2.687$\pm$0.189  &  -1.367$\pm$0.158  & 410 \\
51 &  94.775$\pm$0.166  &  -21.942$\pm$0.177  &  0.085$\pm$0.046  &  0.028$\pm$0.265  &  1.209$\pm$0.261  & 116 \\
52 &  97.885$\pm$0.181  &  37.355$\pm$0.171  &  0.261$\pm$0.121  &  0.558$\pm$0.097  &  -1.905$\pm$0.267  & 86 \\
53 &  70.678$\pm$0.162  &  -75.290$\pm$0.121  &  0.044$\pm$0.034  &  1.935$\pm$0.159  &  0.084$\pm$0.167  & 68 \\
    \hline
  \end{tabular}}
 \end{center}

\end{table*}

\begin{table*}[thp] 
 \footnotesize
 \caption{Similar to Table 5, but for other 30 new OCs.}
 \begin{center}
  \begin{tabular}{cccccccccccccccccccc}
   \hline
NOC ID      &     ra   &    dec   &    $\varpi$  &  $\mu_{\alpha}$~cos$\delta$   &  $\mu_{\delta}$     &  $N_{*}$   \\
          &   [deg]  &   [deg]  &     [mas]    &   [mas/yr]                    &  [mas/yr]         &            \\
\hline
54 &  185.175$\pm$0.162  &  -55.833$\pm$0.161  &  0.098$\pm$0.052  &  -6.135$\pm$0.273  &  0.369$\pm$0.278  & 264 \\
55 &  284.427$\pm$0.170  &  -31.108$\pm$0.170  &  0.040$\pm$0.024  &  -2.650$\pm$0.137  &  -1.391$\pm$0.132  & 598 \\
56 &  72.695$\pm$0.173  &  -73.410$\pm$0.175  &  0.038$\pm$0.023  &  2.036$\pm$0.181  &  -0.094$\pm$0.172  & 428 \\
57 &  281.877$\pm$0.180  &  -31.218$\pm$0.176  &  0.026$\pm$0.014  &  -2.623$\pm$0.131  &  -1.388$\pm$0.107  & 161 \\
58 &  303.148$\pm$0.181  &  52.038$\pm$0.145  &  0.116$\pm$0.060  &  -2.643$\pm$0.197  &  -3.736$\pm$0.251  & 122 \\
59 &  300.908$\pm$0.154  &  21.728$\pm$0.115  &  0.113$\pm$0.053  &  -3.060$\pm$0.280  &  -5.512$\pm$0.279  & 330 \\
60 &  334.451$\pm$0.178  &  48.697$\pm$0.175  &  0.087$\pm$0.046  &  -2.047$\pm$0.281  &  -2.140$\pm$0.282  & 245 \\
61 &  137.442$\pm$0.168  &  -55.768$\pm$0.100  &  0.176$\pm$0.109  &  -3.198$\pm$0.265  &  3.721$\pm$0.127  & 283 \\
62 &  284.360$\pm$0.163  &  -32.088$\pm$0.160  &  0.049$\pm$0.029  &  -2.675$\pm$0.185  &  -1.397$\pm$0.150  & 357 \\
63 &  283.578$\pm$0.186  &  -28.571$\pm$0.176  &  0.036$\pm$0.022  &  -2.702$\pm$0.179  &  -1.441$\pm$0.152  & 212 \\
64 &  11.345$\pm$0.159  &  46.719$\pm$0.163  &  0.172$\pm$0.068  &  -1.287$\pm$0.284  &  -1.115$\pm$0.253  & 40 \\
65 &  108.049$\pm$0.174  &  6.831$\pm$0.177  &  0.197$\pm$0.112  &  -0.216$\pm$0.253  &  -0.403$\pm$0.106  & 265 \\
66 &  311.936$\pm$0.164  &  35.054$\pm$0.158  &  0.192$\pm$0.113  &  -2.564$\pm$0.260  &  -3.991$\pm$0.111  & 414 \\
67 &  338.903$\pm$0.153  &  52.483$\pm$0.165  &  0.099$\pm$0.053  &  -1.985$\pm$0.279  &  -1.605$\pm$0.153  & 182 \\
68 &  286.581$\pm$0.152  &  -32.827$\pm$0.152  &  0.057$\pm$0.035  &  -2.665$\pm$0.169  &  -1.380$\pm$0.161  & 243 \\
69 &  117.597$\pm$0.174  &  -11.972$\pm$0.166  &  0.217$\pm$0.133  &  -1.361$\pm$0.266  &  1.164$\pm$0.124  & 162 \\
70 &  275.711$\pm$0.174  &  4.274$\pm$0.174  &  0.400$\pm$0.067  &  -2.306$\pm$0.274  &  -5.223$\pm$0.280  & 113 \\
71 &  274.396$\pm$0.131  &  5.510$\pm$0.104  &  0.173$\pm$0.109  &  -1.914$\pm$0.245  &  -5.811$\pm$0.108  & 68 \\
72 &  134.480$\pm$0.172  &  -32.839$\pm$0.169  &  0.230$\pm$0.123  &  -2.156$\pm$0.270  &  2.016$\pm$0.127  & 165 \\
73 &  106.496$\pm$0.175  &  -28.987$\pm$0.171  &  0.176$\pm$0.112  &  -0.795$\pm$0.103  &  1.938$\pm$0.259  & 399 \\
74 &  297.140$\pm$0.167  &  12.224$\pm$0.168  &  0.096$\pm$0.040  &  -2.594$\pm$0.264  &  -5.218$\pm$0.240  & 153 \\
75 &  255.013$\pm$0.119  &  -53.853$\pm$0.188  &  0.249$\pm$0.151  &  -3.455$\pm$0.101  &  -5.025$\pm$0.287  & 93 \\
76 &  90.543$\pm$0.163  &  47.532$\pm$0.161  &  0.258$\pm$0.087  &  0.475$\pm$0.115  &  -1.606$\pm$0.278  & 112 \\
77 &  209.252$\pm$0.152  &  -67.055$\pm$0.118  &  0.228$\pm$0.085  &  -6.121$\pm$0.291  &  -2.787$\pm$0.135  & 78 \\
78 &  151.382$\pm$0.150  &  -47.290$\pm$0.166  &  0.220$\pm$0.124  &  -5.018$\pm$0.109  &  3.074$\pm$0.280  & 182 \\
79 &  340.376$\pm$0.131  &  48.294$\pm$0.165  &  0.094$\pm$0.047  &  -1.747$\pm$0.259  &  -1.797$\pm$0.260  & 124 \\
80 &  125.771$\pm$0.150  &  -58.330$\pm$0.169  &  0.120$\pm$0.053  &  -2.535$\pm$0.289  &  3.883$\pm$0.281  & 85 \\
81 &  294.135$\pm$0.154  &  7.751$\pm$0.167  &  0.097$\pm$0.052  &  -2.632$\pm$0.291  &  -5.201$\pm$0.274  & 373 \\
82 &  112.327$\pm$0.163  &  -36.279$\pm$0.159  &  0.099$\pm$0.044  &  -0.910$\pm$0.250  &  2.240$\pm$0.273  & 417 \\
83 &  303.025$\pm$0.162  &  23.490$\pm$0.172  &  0.200$\pm$0.118  &  -2.598$\pm$0.099  &  -4.998$\pm$0.288  & 391 \\
\hline
  \end{tabular}
 \end{center}
\end{table*}

\begin{acknowledgements}
We appreciate the constructive comments of the referee, and thank Mr. Tao Xia and Ms. Yangyang Deng for help.
This work has been supported by Yunnan Academician Workstation of Wang Jingxiu (202005AF150025)
China Manned Space Project (NO.CMS-CSST-2021-A08), and GH project (ghfund202302019167).

This work has made use of data from the European Space Agency (ESA) mission
{\it Gaia} (\url{https://www.cosmos.esa.int/gaia}), processed by the {\it Gaia}
Data Processing and Analysis Consortium (DPAC,
\url{https://www.cosmos.esa.int/web/gaia/dpac/consortium}). Funding for the DPAC
has been provided by national institutions, in particular the institutions
participating in the {\it Gaia} Multilateral Agreement.
\end{acknowledgements}

\section*{Data Availability}
The data underlying this article can be downloaded from Zenodo (DOI 10.5281/zenodo.10154725).

\bibliographystyle{raa}
\bibliography{bibtex}

\label{lastpage}

\end{document}